\documentclass[twocolumn]{aastex702}
\usepackage{amsmath}
\usepackage{lineno}
\linenumbers
\usepackage{subcaption}
\usepackage{ulem}
\usepackage{booktabs}

\newcommand{\soutPC}{\bgroup\markoverwith{\textcolor{cyan}
{\rule[0.5ex]{2pt}{1pt}}}\ULon}

\begin{document}

\title{Gravitational Waves from Population III Black Hole Mergers in Primordial Gas Disks: Constraining Astrophysical Scenarios with the Laser Interferometer Space Antenna and the Einstein Telescope}

\author[orcid=0000-0002-0287-7485]{Maulik Bhatt}
\affiliation{Scuola Normale Superiore, Piazza dei Cavalieri 7, 56126 Pisa, Italy}
\affiliation{Department of Astrophysics, University of Zurich, Winterthurerstrasse 190, CH-8057 Zürich, Switzerland}
\email[show]{maulik.bhatt@sns.it}  

\author[orcid=0000-0002-7078-2074]{Lucio Mayer}
\affiliation{Department of Astrophysics, University of Zurich, Winterthurerstrasse 190, CH-8057 Zürich, Switzerland}
\email[]{lucio.mayer@uzh.ch}

\author[orcid=0000-0002-1786-963X]{Pedro R. Capelo}
\affiliation{Department of Astrophysics, University of Zurich, Winterthurerstrasse 190, CH-8057 Zürich, Switzerland}
\email[]{pedro.capelo@uzh.ch}

\author[orcid=0000-0002-7200-8293]{Simona Gallerani}
\affiliation{Scuola Normale Superiore, Piazza dei Cavalieri 7, 56126 Pisa, Italy}
\email[]{simona.gallerani@sns.it}

\author[orcid=0000-0002-7129-5761]{Andrea Pallottini}
\affiliation{Dipartimento di Fisica `Enrico Fermi', Università di Pisa, Largo Bruno Pontecorvo 3, 56127 Pisa, Italy}
\email[]{andrea.pallottini@unipi.it}

\begin{abstract}
The first generation of metal-free Population III (Pop~III) stars provides a possible formation channel for the black hole (BH) seeds in the early Universe. Their remnants may grow through gas accretion and mergers, producing gravitational wave (GW) signals detectable by future observatories. We develop a semi-analytical model of fragmentation in a primordial self-gravitating gas disk within minihaloes at $z > 15$, motivated by high-resolution cosmological simulations. The resulting Pop~III stars form around a central primary, some leaving massive BH remnants that migrate towards the central BH through dynamical friction from gas and dark matter, while the central BH grows through accretion and mergers. For a $10^6~{\rm M_\odot}$ halo at $z=20$, we explore how the initial mass function (IMF) and BH accretion efficiency affect detectability with the Laser Interferometer Space Antenna (LISA) and the Einstein Telescope (ET). We find that both detectors can potentially reveal mergers of Pop III remnant BHs, but favor different astrophysical conditions. LISA pinpoints more massive central BHs ($\sim10^3$--$10^4 {\rm M_\odot}$), grown rapidly through mutual mergers alongside efficient gas accretion ($f_{\rm Edd}=1$), and a relatively top-heavy IMF ($\alpha=0.8$). In contrast, detectability with ET favors lower-mass BHs, with less efficient accretion ($f_{\rm Edd}=0.01$) and a steeper IMF ($\alpha=1.5$). These results highlight the potential of LISA and ET to probe Pop~III BH formation, growth, and mergers across different regions of parameter spaces, jointly constraining Pop~III star formation and early BH seed evolution.
\end{abstract}

\keywords{}


\section{Introduction}

Future gravitational wave (GW) observatories, including the Laser Interferometer Space Antenna \citep[LISA;][]{Colpi2024}, Einstein Telescope \citep[ET;][]{Abac2026}, Cosmic Explorer \citep[CE;][]{Hall2022}, TianQin \citep{Luo2016, Mei2021}, and Taiji \citep{Hu2017}, will provide new opportunities to detect black hole (BH) mergers at very high redshifts. Their complementary sensitivities across different GW frequency bands will allow us to probe a broad range of BH masses and investigate the formation and growth of the first BHs in the early Universe.

The discovery of luminous quasars hosting supermassive BHs (SMBHs) with masses $\gtrsim10^9~{\rm M_\odot}$ at redshift $z\gtrsim6$--$7$ \citep{Fan2001, fan2006, Banados2016}, together with recent observations of accreting BHs at even higher redshifts by the James Webb Space Telescope \citep[JWST;][]{greene2023, Harikane2023, Maiolino2023b, Kokorev2024, Maiolino2024, Kocevski2025, Taylor2026}, challenges our understanding of early BH formation and growth. Their rapid assembly requires sufficiently early BH seed formation followed by efficient growth through gas accretion and/or mergers \citep{Inayoshi2020}. However, their initial masses, formation epochs, and subsequent growth histories remain uncertain. GW observations offer an independent probe of these properties through the masses, mass ratios, and merger redshifts of BH binaries, potentially distinguishing different BH seeding scenarios \citep{Sesana2007, volonteri2021}.

One possible formation channel for early BH seeds is the collapse of Population III (Pop~III) stars, which form from metal-free primordial gas in the first dark matter (DM) minihaloes at $z\sim20$ and above \citep{Abel2002,Yoshida2006,Hirano2014}. Numerical simulations show that primordial protostellar discs can fragment into multiple stars, with gravitational torques driving their migration \citep{Clark2011b,Greif2012}. These processes produce Pop~III stars over a broad mass range, some of which leave behind BH remnants \citep{Hirano2014,Klessen2023}.

Whether these remnants subsequently grow and merge remains uncertain. Previous studies have explored primordial binaries and BH mergers through dynamical capture \citep{Stacy2013,Liu2020}. Alternatively, BH remnants may migrate towards a central BH through dynamical friction against dark matter \citep{Chandrasekhar1943} and gas \citep{Ostriker1999}. Related mechanisms have been investigated in gas-rich high-redshift protogalaxies \citep{Tagawa2015,Ryu2016,Safarzadeh2020}, progenitors of present-day early-type galaxies \citep{Boco2020}, and nuclear star clusters \citep{Liu2021b}. However, their efficiency in primordial minihaloes at $z>15$ remains poorly constrained. Remnant migration depends on disk properties and remnant distributions, while BH growth depends on gas supply and feedback \citep{Smith2018,Gordon2024,Mehta2026}.

In this work, we develop a semi-analytical framework connecting primordial disc fragmentation, Pop~III star formation, and stellar remnant production to their subsequent orbital decay, accretion, and mergers with a central BH. We investigate whether dynamical friction can bring these remnants towards the central BH within the isolated-halo evolution time-scale and, assuming prompt coalescence, whether the resulting GW signals could be detectable by LISA or ET. By exploring the disc properties, Pop~III initial mass function (IMF), and BH accretion efficiency, we identify the astrophysical conditions favouring detection in each frequency band. Rather than predicting cosmological merger rates, our goal is to demonstrate the possibility of detectable events and investigate the complementary potential of LISA and ET to probe early BH formation and growth.

\section{Methodology}\label{sec:method}

In this section, we briefly describe the model for Pop~III star formation in a primordial disk, followed by the subsequent inspiral of the remnants and the resulting GW signatures from their mergers. 

We consider a mini-halo of virial mass $M_{\rm vir}=10^6\,{\rm M_\odot}$ at $z=20$, motivated by cosmological simulations identifying $\sim10^5$--$10^6\,{\rm M_\odot}$ haloes at $z\sim20$--30 as typical sites of Pop~III star formation through molecular hydrogen cooling \citep{Abel2002,Bromm2004,Bromm2013,Stacy2010}.

We first characterize the host DM halo and its associated baryonic disk, which provide the initial conditions for the subsequent fragmentation and star formation. We calculate the virial radius from the virial mass and redshift assuming a spherical-collapse model. The virial radius is given by

\begin{equation}
R_{\rm vir} = \left[ \frac{2 G M_{\rm vir}} {\Delta_{\rm vir}(z)H^2(z)}
\right]^{1/3},
\label{eq:rvir_z}
\end{equation}

\noindent where $G$ is the gravitational constant and  $H(z)=H_0[\Omega_{\rm m,0}(1+z)^3+\Omega_{\Lambda,0}]^{1/2}$ is the Hubble parameter for a spatially flat $\Lambda$CDM cosmology\footnote{We adopted the cosmological parameters measured by \cite{Planck2020}: $\Omega_{\rm m,0} = 0.3153$, $\Omega_{\rm b,0} = 0.0493$, $\Omega_{\Lambda,0} = 0.6847$, $h = 0.6736$, $n_{\rm s} = 0.9649$, and $\sigma_{8} = 0.8111$.}. Here, $H_0=100~h\,\mathrm{km\,s^{-1}\,Mpc^{-1}}$ is the present-day Hubble constant, whereas $\Omega_{\rm m,0}$ and $\Omega_{\Lambda,0}$ denote the present-day matter and dark-energy density parameters, respectively.
We adopt the redshift-dependent virial overdensity from \citet{Bryan1998},
$\Delta_{\rm vir}(z)=18\pi^2+82x-39x^2,$
where $x=\Omega_{\rm m}(z)-1$ and

\begin{equation}
    \Omega_{\rm m}(z)= \frac{\Omega_{\rm m,0}(1+z)^3} {\Omega_{\rm m,0}(1+z)^3+\Omega_{\Lambda,0}}.
\end{equation}

Here, $\Delta_{\rm vir}(z)$ is defined relative to the critical density of the Universe, $\rho_{\rm cr,0}=2.775\times10^{11} h^2~\mathrm{M}_\odot~{\rm Mpc}^{-3}$. 
At the high redshifts considered in this work, $z\gtrsim10$, the Universe is matter-dominated, and the virial overdensity approaches the Einstein–de Sitter value, $\Delta_{\rm vir}\simeq18\pi^2\simeq178$. The circular velocity at the virial radius is given by

\begin{equation}
V_{\rm vir} = \sqrt{\frac{G M_{\rm vir}}{R_{\rm vir}}}.
\end{equation}

Thus, for a given halo virial mass and redshift, the virial radius and circular velocity are determined self-consistently and set the characteristic dynamical scale of the halo.

\subsection{Disk model}

We assume that a fraction $f_{\rm d}$ of the halo's
baryonic content settles into a thin, rotationally
supported disk, with a total mass

\begin{equation}
M_{\rm d}=f_{\rm b}f_{\rm d}M_{\rm vir},
\label{eq:Md}
\end{equation}

\noindent where $f_{\rm b}=\Omega_{\rm b,0}/\Omega_{\rm m,0}$ is the cosmic baryon fraction. Observations of high-redshift galaxies reveal substantial gas reservoirs and baryon-dominated inner disks \citep{Tacconi2013, Genzel2017}, although they do not directly constrain primordial mini-haloes. We therefore explore $f_{\rm d}=0.05$--$0.5$ to account for uncertainties in the efficiency of baryonic disk assembly and the gas available for Pop~III star formation.

Following the angular-momentum argument of \citet{Mo1998}, we assume an exponential disk with a characteristic scale length,

\begin{equation}
R_{\rm d}=
\frac{1}{\sqrt{2}}
\frac{j_{\rm d}}{m_{\rm d}}
\lambda R_{\rm vir},
\label{eq:Rd}
\end{equation}

\noindent appropriate for a disk embedded in a singular isothermal halo, where $j_{\rm d}/m_{\rm d}$ represents the disk-to-halo specific angular momentum ratio, which we assume to be unity, and $\lambda$ is the halo spin parameter \citep{Bullock2001}. Motivated by cosmological simulations, we explore $\lambda=0.01$--$0.05$ \citep{Yoshida2003,Bett2007,Munoz-Cuartas2011,Sasaki2014, Kohandel2025}. This range is also broadly consistent with observational estimates for star-forming galaxies at $z\sim1$--$3$, which suggest $\lambda\simeq0.037$ \citep{Burkert2016}.

The disk surface density is described by an exponential profile,

\begin{equation}
\Sigma(R)=\Sigma_0 e^{-R/R_{\rm d}},
\qquad
\Sigma_0=\frac{M_{\rm d}}{2\pi R_{\rm d}^2},
\label{eq:sigma}
\end{equation}

\noindent where $\Sigma_0$ is the central surface density, determined by the disk mass and scale length. 



Cosmological simulations show that $\rm H_2$ line cooling lowers primordial gas temperatures to $T\simeq200$ K at $n_{\rm H}\sim10^4\,{\rm cm^{-3}}$ \citep{Abel2002, Bromm2002, Yoshida2006}, while subsequent collapse and dynamical heating can raise the temperature \citep{Yoshida2006, Greif2012}. We therefore explore $T_{\rm disk}=200$--$1000$ K to assess its effects on disk stability, fragmentation, and Pop~III star formation.


This disk model provides the radial gas distribution used to determine the gravitational stability and fragmentation of the primordial disk.

\subsection{Rotation curve and epicyclic frequency}\label{sec:rotation_curve}

The gas disk orbits in the gravitational potential of the DM halo, and its rotation curve is set by the DM profile alone. We model the halo with an NFW profile \citep{Navarro1997} of concentration $c_{\rm NFW}=4$ \footnote{Extrapolating the concentration relation of \citet{Angel2016} to $M_{\rm vir}=10^6\,{\rm M_\odot}$ at $z=20$ gives $c_{\rm vir}\simeq3.4$, close to our fiducial value. However, this estimate lies outside the simulation's calibrated range and should be treated as indicative.} and scale radius $r_{\rm s}=R_{\rm vir}/c_{\rm NFW}$. The DM mass enclosed within a radius $R$ is

\begin{equation} 
\begin{aligned} 
M_{\rm DM}(<R) &= M_{\rm DM,vir}\,\frac{f(u)}{f(c_{\rm NFW})}, \\ f(u) &= \ln(1+u)-\frac{u}{1+u}, 
\end{aligned} 
\label{eq:nfw_mass} 
\end{equation}

\noindent where $u=R/r_{\rm s}$ and $M_{\rm DM,vir}=(1-f_{\rm b})\,M_{\rm vir}$. The circular velocity and angular frequency are then, respectively,

\begin{equation}
  V_{\rm c}(R)=\sqrt{\frac{G\,M_{\rm DM}(<R)}{R}},
  \qquad
  \Omega(R)=\frac{V_{\rm c}(R)}{R}.
  \label{eq:vc}
\end{equation}

The circular velocity $V_{\rm c}(R)$ sets the orbital velocity of the clumps and hence enters the dynamical-friction and gas-drag migration time-scales. The epicyclic frequency 
$\kappa$, defined by $\kappa^2=(2\Omega/R)\,\mathrm{d}(R^2\Omega)/\mathrm{d}R$, becomes, for the NFW profile,

\begin{equation} 
\kappa(R) = \Omega(R) \sqrt{1+\frac{u^2}{(1+u)^2f(u)}}.
\label{eq:kappa} 
\end{equation}

We assume that the rotation curve is determined solely by the DM halo, neglecting the gravitational contribution of the gas disk and stars, which may lead to an underestimation of the circular velocity in the inner disk. We also neglect the adiabatic contraction of the DM halo due to baryons and assume a static, spherical halo with a fixed concentration.

\subsection{Fragmentation and clump formation}

The gravitational stability of the primordial disk is determined by the balance between self-gravity, gas pressure, and rotational support. We characterize this using the Toomre parameter \citep{Toomre1964},

\begin{equation}
Q(R) = \frac{\kappa(R)c_{\rm s}}{\pi G\Sigma(R)},
\label{eq:Q}
\end{equation}

\noindent where we opt for an isothermal sound speed, $c_{\rm s} = \sqrt{\frac{\,k_{\rm B}\,T}{\mu\,m_{\rm H}}}$,
where $k_{\rm B}$ is the Boltzmann constant, $m_{\rm H}$ is the mass of the hydrogen atom, and $\mu = 1.22$ is the mean molecular weight of neutral atomic gas (H and He). For $T=300$~K, characteristic of H$_2$-cooled primordial gas in Pop~III mini-haloes, this yields $c_{\rm s}\simeq1.42\,{\rm km\ s^{-1}}$. 

For a thin gaseous disk, local axisymmetric instability occurs at $Q<1$, while simulations of locally isothermal disks show that fragmentation can occur at $Q\lesssim1.4$ due to non-axisymmetric instabilities \citep{Johnson2003, Mayer2004, Durisen2007}. Assuming sufficiently efficient cooling, we adopt $Q_{\rm crit}=1.4$ as the fragmentation threshold, following \citet{Derdzinski2023}. 

In the gravitationally unstable region, the characteristic scale of fragmentation is estimated from the most unstable wavelength and its corresponding Toomre mass \citep{Toomre1964, Derdzinski2023}:

\begin{align}
\lambda_{\rm T}(R) &= \frac{2 c_{\rm s}^2}{G \Sigma(R)},
\label{eq:lambdaT} \\ 
M_{\rm T}(R)   &= \pi \Sigma(R)\left(\frac{\lambda_{\rm T}}{2}\right)^2
= \frac{\pi c_{\rm s}^4}{G^2 \Sigma(R)} .
\label{eq:MT}
\end{align}

Here, $M_{\rm T}(R)$ represents the characteristic gas mass associated with gravitationally unstable perturbations. Following this approximation, we use $M_{\rm T}(R)$ to set the local maximum mass available for an individual stellar fragment.

Assuming that the vertical structure of the disk is supported by gas pressure against its own self-gravity, the disk scale height is

\begin{equation}\label{eq:scale_height}
h(R)=\frac{c_{\rm s}^2}{\pi G\Sigma(R)}.
\end{equation}

The corresponding midplane gas density and free-fall time are, respectively

\begin{equation}
\rho(R)=\frac{\Sigma(R)}{2h(R)},
\qquad
t_{\rm ff}(R)=\sqrt{\frac{3\pi}{32 G\rho(R)}}.
\label{eq:rho}
\end{equation}

The free-fall time provides an estimate of the characteristic time-scale over which gravitationally bound gas fragments collapse.

We identify the gravitationally unstable region as the radial range where $Q(R)<Q_{\rm crit}$, bounded by $[R_{\rm in},R_{\rm disk}]$. Assuming that the gas in this region fragments into clumps with characteristic masses given by the local Toomre mass, the expected number of fragments per unit radius is estimated as

\begin{table*}[t]
  \centering
  \caption{Fates and remnant masses adopted for zero-metallicity stars
  \citep{Heger2002, Heger2003}, while the remnant masses are simplified prescriptions. $M$ is the zero-age main-sequence mass.}
  \label{tab:remnants}
  \begin{tabular}{lll c}
    \hline
    Mass range $[\rm M_\odot]$ & Fate & Remnant mass $[\rm M_\odot]$ & BH? \\
    \hline
    $0.8 \le M < 8$  & white dwarf (WD)                  & $0.109\,M+0.394$ & no  \\
    $8 \le M < 25$   & neutron star (core-collapse SN)           & $1.4$            & no  \\
    $25 \le M < 40$  & fallback BH           & $0.5\,M$         & yes \\
    $40 \le M < 100$ & direct-collapse BH    & $0.9\,M$         & yes \\
    $100 \le M < 140$& pulsational pair-instability SN $\to$ BH      & $0.5\,M$         & yes \\
    $140 \le M < 260$& pair-instability SN           & $0$              & no  \\
    $M \ge 260$      & direct-collapse BH     & $0.9\,M$         & yes \\
    \hline
  \end{tabular}
\end{table*}

\begin{equation}
\frac{{\rm d}N}{{\rm d}R}
=\frac{2\pi R\Sigma(R)}{M_{\rm T}(R)}.
\label{eq:dNdR}
\end{equation}

Integrating over the unstable region gives the total geometric fragment count,

\begin{equation}
N_{\rm geom}=
\int_{R_{\rm in}}^{R_{\rm disk}}
\frac{2\pi R\Sigma(R)}{M_{\rm T}(R)} {\rm d}R.
\label{eq:N}
\end{equation}

This represents the number of fragments expected if the unstable gas is divided into clumps of the local Toomre mass, before accounting for subsequent accretion, mergers, or other processes that may modify the final stellar population. In reality, clumps would form with a spectrum of masses and not all at the Toomre mass, but numerical simulations suggest that low-mass clumps are either tidally disrupted or accreted by the more massive clumps on a local orbital time-scale, hence our assumption that a fragmenting region is turned into Toomre-mass fragments is reasonable. \cite{Boley2010} introduced a modified Toomre mass that better captures the fact that fragmentation in 3D global simulations of self-gravitating disks is happening in the non-axisymmetric flow (along spiral arms) rather than in the axisymmetric conditions postulated by the Toomre instability theory. Such modified fragmentation mass is smaller by factors of a few compared to the conventional Toomre mass, which has been verified in disks ranging from protoplanetary disks to the massive galactic gas disks of cosmic noon galaxies \citep{Tamburello2015} and high-redshift compact disks forming star clusters \citep{Mayer2025}. However, in this work, as we explain in the next section, the Toomre mass is essentially used to provide a ceiling on the mass function of stars, while the mass spectrum of stars is initialized based on the initial mass function (IMF)  suggested from numerical simulations of Pop~III star formation. Therefore, in order to keep the model as simple as possible, we do not introduce any additional correction on the initial mass of fragments.

\subsection{Initial mass function}

The IMF of Pop~III stars remains uncertain, as fragmentation, accretion, and dynamical interactions in primordial gas can produce a broad distribution of stellar masses \citep{Clark2011, Hirano2014, Susa2014, Stacy2016}. 
Motivated by numerical simulations \citep{Susa2014, Stacy2016, Latif2022, Pustovoit2026} and recent observational indications \citep{Rusta2026} favoring a top-heavy Pop~III IMF, we adopt a single power-law IMF,

\begin{equation}
\frac{{\rm d}N}{{\rm d}M} \propto M^{-\alpha},
\qquad M \in [M_{\rm min}, M_{\rm max}],
\label{eq:imf}
\end{equation}

\noindent where $M_{\rm min}$ and $M_{\rm max}$ are the stellar mass limits and $\alpha$ is the IMF slope, for which we explore the range $\alpha=0.8$--$1.5$, corresponding to shallower slopes than the classical Salpeter value of $\alpha=2.35$ \citep{Salpeter1955}. 



Radiation-hydrodynamic and cosmological simulations suggest that Pop~III stars can reach $\sim10^3$--$10^4\,{\rm M_\odot}$ in massive haloes, with a steep decline in their abundance at higher masses \citep{Toyouchi2023,Hirano2026}. Stellar evolution models also support the existence of rotating Pop~III stars in this mass range \citep{Nandal2025}. We therefore adopt $M_{\rm min}=1\,{\rm M_\odot}$ and $M_{\rm max,global}=10^4\,{\rm M_\odot}$.

The local maximum stellar mass is limited by the Toomre mass, $M_{\rm max}(R)=\min[M_{\rm max,global}, \varepsilon_{\rm clump}M_{\rm T}(R)]$, where $\varepsilon_{\rm clump}$ is the fraction of the local collapse mass converted into a star. To account for uncertainties arising from fragmentation, accretion, and dynamical interactions \citep{Clark2011,Greif2012,Hirano2017}, we explore $\varepsilon_{\rm clump}=0.1$--$1$.

The total stellar mass is constrained by the gas available in the
gravitationally unstable region,

\begin{equation}
M_{\star,\rm budget}=\varepsilon_{\rm local,SF}
M_{\rm gas}(Q<Q_{\rm crit}),
\end{equation}

\noindent where $\varepsilon_{\rm local,SF}$ is the local star formation efficiency, regulated by fragmentation, accretion, and radiative feedback \citep{Hirano2014,Stacy2016, Reinoso2025}. We explore $\varepsilon_{\rm local,SF}=0.01$--$0.1$ and sample stellar masses from the IMF without exceeding this budget.


\subsection{Lifetimes, fates, and remnant masses}

Following disk fragmentation, we assign each bound clump a protostellar evolution time $t_{\rm proto}$, drawn uniformly from $0$--$0.1$ Myr. This phenomenological delay is motivated by Kelvin--Helmholtz time-scales of a few $\times10^{-2}$ Myr and the accretion-dependent approach to the zero-age main sequence (ZAMS) \citep{Hirano2017}, without explicitly modelling protostellar accretion.


We estimate the main-sequence lifetimes using a polynomial fit to zero-metallicity stellar evolution models \citep{Schaerer2002},

\begin{equation}
\begin{aligned}
\log\!\left(\frac{t_\star}{\rm yr}\right)
&=9.785-3.759\,x+1.413\,x^2-0.186\,x^3,\\
x&\equiv\log\!\left(\frac{M}{\rm M_\odot}\right).
\end{aligned}
\label{eq:lifetime}
\end{equation}

The fit is calibrated for $5$--$500\,{\rm M_\odot}$, with lifetimes approaching $\sim2$ Myr at the upper end, and extrapolated to $1$--$1000\,{\rm M_\odot}$ in our model. For $M>1000\,{\rm M_\odot}$, we use the tabulated lifetimes of \citet{Nandal2025} for $v/v_{\rm crit}=0.1$ (third column of their Table~1), extending to $10^4\,{\rm M_\odot}$ with lifetimes of $\sim1.15$--$1.85$ Myr. Neglecting post-main-sequence evolution, we approximate the stellar death time as $t_{\rm death}=t_{\rm proto}+t_\star$, measured from clump formation.

We adopt approximate zero-metallicity stellar fate boundaries based on \citet{Heger2002,Heger2003}, with pair-instability supernovae leaving no remnant. The remnant masses are assigned using the simplified prescriptions listed in Table~\ref{tab:remnants}.


\subsection{Spatial distribution}

Simulations suggest that the first Pop~III protostar forms near the center of the collapsing gas, followed by additional protostars forming through disk fragmentation at larger radii \citep{Clark2011b, Haemmerle2020}. Motivated by this picture, we assume that the progenitor of the most massive BH remnant forms at the disk centre and subsequently produces the primary BH. The remaining BH remnants are treated as satellites that undergo orbital decay through dynamical friction.

The initial stellar radii are sampled from the fragmentation profile (Equation~\ref{eq:dNdR}), normalised over $R\in[R_{\rm in},R_{\rm disk}]$, where $Q<Q_{\rm crit}$. This favors regions with higher gas surface densities and smaller fragmentation masses. Assuming an axisymmetric disk, azimuthal positions are drawn uniformly from $[0,2\pi)$, while vertical positions follow a Gaussian distribution centered on the midplane with standard deviation $h(R)$  (Equation~\ref{eq:scale_height}).

Simulations also predict Pop~III binaries and multiple systems, with binary fractions of $\sim35\%$, separations up to several thousand AU, and orbital periods of $\sim30$--$30{,}000$ yr \citep{Stacy2013, Stacy2016}. Although such systems may influence the subsequent remnant dynamics and mergers, we do not explicitly model binary evolution or interactions amongst the satellites.

\subsection{Migration, accretion, and mergers}\label{sec:migration}

Following Pop~III BH remnant formation, we model orbital decay due to dynamical friction from the DM halo and the disc gas. Assuming circular orbits, the angular momentum of a remnant of mass $m_i$ at radius $r_{i}$ is $L_i=m_i r_i v_i$, where

\begin{equation}
v_i^2(r)=v_{i,\rm c}^2(r)+\frac{GM_{1,\rm ini}}{r_i},
\label{eq:vorb}
\end{equation}

\noindent with $v_{i,\rm c}$ representing the background potential
and $M_{1_,ini}$ the initial primary BH mass. The primary's gravity becomes important where $GM_1/r_i\gtrsim v_{i,\rm c}^2$. Assuming drag opposes orbital motion,
${\rm d}L_i/{\rm d}t=-r_iF_i$ yields

\begin{equation}
\frac{{\rm d}r_i}{{\rm d}t}
=-\frac{r_iF_i}
{m_i\left(v_i+r_i\,{\rm d}v_i/{\rm d}r\right)}.
\label{eq:drdt}
\end{equation}

We combine the Chandrasekhar dynamical friction from DM
\citep{Chandrasekhar1943,Binney2008} with gaseous drag
\citep{Ostriker1999},

\begin{equation}
F=
\frac{4\pi G^2m_i^2}{v_i^2}
\left[
\frac{\rho_{\rm halo}\ln\Lambda_i}{\mathcal{C}}
+\rho_{\rm gas}I_i(\mathcal{M}_i)
\right],
\label{eq:force}
\end{equation}

\noindent where the local NFW halo density is

\begin{equation}
\rho_{\rm halo}(r)=
\frac{\rho_{\rm s}}
{(r/r_{\rm s})(1+r/r_{\rm s})^2},
\qquad
\rho_{\rm s}=
\frac{M_{\rm DM,vir}}{4\pi r_{\rm s}^3f(c_{\rm NFW})},
\label{eq:rhonfw}
\end{equation}

\noindent with $\rho_{\rm s}$ the characteristic density. We adopt $\mathcal C=1/0.428\simeq2.34$, corresponding to the Chandrasekhar velocity-distribution correction for circular orbits in an isothermal halo \citep{Binney2008}, and apply this value approximately to the NFW halo. The gas density is 

\begin{equation}
\rho_{\rm gas}(r,z)=
\frac{\Sigma(r)}{2h(r)}
\exp\left(-\frac{|z|}{h(r)}\right),
\label{eq:rhogas}
\end{equation}

\noindent with each remnant retaining its initial vertical position $z$ during orbital decay. We adopt a position- and mass-dependent Coulomb logarithm, $\ln\Lambda_{i}=\ln\max[R_{\rm disk}/b_{{\rm min},i},2]$, where $b_{{\rm min},i}=2Gm_i/(v_i^2+c_{\rm s}^2)$ is the gravitational focusing scale and $R_{\rm disk}$ sets the maximum interaction scale. We update $\ln\Lambda_i$ along each orbit and apply it to both DM and gaseous friction.

For the gaseous component, we approximate the velocity of the remnant relative to the surrounding gas by $v_{\rm c}$, i.e. we treat the gas as static and neglect disk rotation. The corresponding Mach number is $\mathcal{M}_i=v_i/c_{\rm s}$, where $c_{\rm s}$ is the thermal sound speed of the gas. Following \citet{Ostriker1999}, the dimensionless drag factor is

\begin{equation}
I_{i,\rm raw}(\mathcal{M}_i)=
\begin{cases}
\displaystyle
\frac{1}{2}\ln\left(\frac{1+\mathcal{M}_i}{1-\mathcal{M}_i}\right)-\mathcal{M}_i,
& \mathcal{M}_i\leq0.99,\\[1.2ex]
\displaystyle
\ln\Lambda_i+\frac{1}{2}\ln\left(1-\mathcal{M}_i^{-2}\right),
& \mathcal{M}_i\geq1.01.
\end{cases}
\label{eq:ostriker}
\end{equation}

We regularize the transonic regime by linearly interpolating between the two branches over $0.99<\mathcal{M}_i<1.01$. Since the supersonic expression can become negative near $\mathcal{M}_i=1$ for small Coulomb logarithms, we impose $I(i, \mathcal{M}_i)=\max[0,I_{i, \rm raw}(\mathcal{M}_i)]$, where $I_{i, \rm raw}$ includes the transonic interpolation. This numerical regularization prevents unphysical negative drag forces. Our treatment neglects modifications to the gravitational wake arising from orbital curvature and the detailed gas motion in a rotating disk \citep{Kim2007,Escala2004,Chapon2013}.

The corresponding dynamical-friction sinking time is

\begin{equation}
t_{{\rm fri},i}=
\int_{r_{{\rm f},i}}^{r_i}
\frac{
(v_i+r\,dv_i/dr)\,v_i^2
}{
4\pi G^2m_i r
\left[
\rho_{\rm halo}\ln\Lambda_i/\mathcal C+
\rho_{\rm gas}I_i(\mathcal M_i)
\right]
}\,dr.
\label{eq:tfri}
\end{equation}

We integrate Equation~\eqref{eq:tfri} from the initial remnant radius $r_i$ to the Bondi radius of the primary BH, $r_{{\rm f},i}$, inside which the primary dominates the gas dynamics. Once stars and BHs have formed, we assume that the disk self-regulates to a marginally stable gravitoturbulent state with Toomre parameter $Q_{\rm target}=1.5$. We therefore evaluate the Bondi radius with the effective sound speed

\begin{equation}
c_{{\rm eff},i}=\max\left[c_{\rm s},\,
\frac{Q_{\rm target}\pi G\Sigma(r_i)}{\kappa(r_i)}\right],
\label{eq:ceff}
\end{equation}

\noindent computed at the initial radius of each remnant and kept fixed along its orbit, so that

\begin{equation}
r_{{\rm f},i}=\min\left(\frac{2GM_{1,\rm ini}}{c_{{\rm eff},i}^2},\,0.5R_{\rm disk}\right),
\label{eq:rf}
\end{equation}

\noindent Since $c_{{\rm eff},i}\geq c_{\rm s}$, this endpoint lies inside the thermal Bondi radius.\footnote{Replacing $c_{{\rm eff},i}$ by $c_{\rm s}$ changes the number of detectable events by less than 5 per cent.} We assume prompt coalescence once the remnant reaches $r_{{\rm f},i}$, neglecting the additional delay associated with binary hardening and GW inspiral. 

We neglect stellar migration before remnant formation and mass growth of secondary BHs during inspiral, approximating
the merger time as $t_{{\rm merge},i}=t_{{\rm death},i}+t_{{\rm fri},i}$. Since sinking times increase with initial radius and decrease with remnant mass, more massive BHs closer to the disc center generally merge earlier.

The primary BH begins accreting gas at $t_{{\rm death},1}$, following the collapse of its progenitor star. The accretion efficiency of Pop~III remnants remains uncertain, with cosmological simulations finding negligible growth ($\sim10^{-5}$ in fractional mass in simulations spanning $\sim300$ Myr) for most remnants \citep{Smith2018}, while others predict substantial growth in dense primordial gas, reaching $\sim10^3\,{\rm M_\odot}$ within $\sim1$ Myr \citep{Gordon2024} or BH masses of $\gtrsim10^4\,{\rm M_\odot}$ within a few Myr \citep{Mehta2026}. The former rapid-growth calculation neglects radiative feedback and subsequent star formation, whereas the latter includes SN and BH thermal feedback. Motivated by the uncertain efficiency of Pop~III BH accretion, we assume that only a fraction $f_{\rm gas,BH}$ of the gas within the evolving BH capture radius is available for accretion. We adopt $f_{\rm gas,BH}=0.01$ and explore values up to $0.1$ to account for variations in the available gas supply.

The primary BH grows through gas accretion and mergers. Starting from its formation at the death of the primary star, it accretes at

\begin{equation}
\dot{M}_{\rm BH}=
\min\left[
f_{\rm rem}\dot{M}_{\rm B},\,
\frac{f_{\rm Edd}M_{\rm BH}}{t_{\rm Sal}}
\right],
\label{eq:bh_accretion}
\end{equation}

\noindent where $\dot{M}_{\rm B}=4\pi\lambda_{\rm acc}G^2M_{\rm BH}^2\rho_0/c_{\rm s}^3$ is the Bondi rate, $\lambda_{\rm acc}=1$ is a dimensionless accretion efficiency, and $\rho_0=\pi G\Sigma_0^2/(2c_{\rm s}^2)$ is the central midplane gas density of the disk, with $c_{\rm s}$ the thermal sound speed. The second term is the Eddington-limited rate, where $f_{\rm Edd}$ is the Eddington ratio and $t_{\rm Sal}=t_{\rm Edd}\,\epsilon_{\rm rad}/(1-\epsilon_{\rm rad})\simeq50$~Myr is the Salpeter e-folding time for a radiative efficiency $\epsilon_{\rm rad}=0.1$, with $t_{\rm Edd}\simeq450$~Myr.

To account for the finite gas supply, we assume that only a fraction $f_{\rm gas,BH}$ of the gas remaining within the BH capture radius is available for accretion:

\begin{equation}
M_{\rm local}=f_{\rm gas,BH}
\left[M_{\rm gas}(<R_{\rm B})-M_\star(<R_{\rm B})\right],
\label{eq:mlocal}
\end{equation}

\noindent where
$R_{\rm B}=\min(2GM_{\rm BH}/c_{\rm s}^2,R_{\rm disk})$ is the effective BH capture radius. Here, $M_{\rm gas}(<R_{\rm B})$ is the enclosed gas mass before star formation, while $M_\star(<R_{\rm B})$ is estimated by distributing the total stellar mass across the fragmenting region in proportion to the initial gas mass. As the BH grows, we update $R_{\rm B}$ and $M_{\rm local}$, accounting for gas depletion through $f_{\rm rem}=\max(0,1-M_{\rm acc}/M_{\rm local})$, where $M_{\rm acc}$ is the cumulative gas mass accreted by the primary. Accretion is halted when the local reservoir is exhausted.

We assume that the primordial disk remains intact during inspiral and neglect interactions amongst satellite remnants. However, gas-driven migration and dynamical interactions may promote BH binary formation and alter their orbits \citep{Liu2020, Liu2021}. Gas-assisted capture in active galactic nucleus disks provides a related mechanism, although its relevance to primordial disks remains uncertain \citep{Rozner2023, DeLaurentiis2023, Rowan2023}. Changes in orbital eccentricity may also affect the GW signals \citep{Wu2026}. These effects are neglected in our model.

\subsection{Gravitational wave signal}

\begin{table}
\centering
\caption{Combinations of free parameters yielding detectable events (SNR~$\ge8$) for ET (EDEs, ET-detectable events) and LISA (LDEs, LISA-detectable events) in a single halo with $M_{\rm vir}=10^6\rm M_\odot$ at $z=20$ (1000 random draws for each parameter combination).}
\begin{tabular}{lccc}
\hline
 & Favorable & Favorable \\
Explored Parameters & ET values & LISA values \\
\hline
Disk baryon fraction, $f_{\rm disk}$       & $0.3$  & $0.5$  \\
Halo spin, $\lambda$                       & $0.03$ & $0.02$ \\
Gas temperature, $T_{\rm disk}$ [K]                   & $200$  & $300$  \\
IMF slope, $\alpha$                        & $1.5$  & $0.8$  \\
Clump efficiency, $\epsilon_{\rm clump}$   & $0.1$  & $1.0$  \\
local SF efficiency, $\epsilon_{\rm local,SF}$ & $0.01$ & $0.1$ \\
Eddington ratio, $f_{\rm Edd}$             & $0.01$  & $1.0$ \\
gas supply primary BH, $f_{\rm gas,BH}$    & $0.01$ & $0.01$ \\
\hline
\hline
Fixed parameter \\ 
\hline 
Toomre parameter, $Q_{\rm crit}$ & 1.4 & 1.4 \\
radiative efficiency, $\epsilon_{\rm r}$ & 0.1 & 0.1 \\
accretion efficiency, $\lambda_{\rm acc}$ & 1 & 1 \\
\hline
\end{tabular}
\label{tab:best_params}
\end{table}

\begin{figure*}[t]
    \centering
    \begin{subfigure}[b]{0.325\textwidth}
        \centering
        \includegraphics[width=\linewidth]{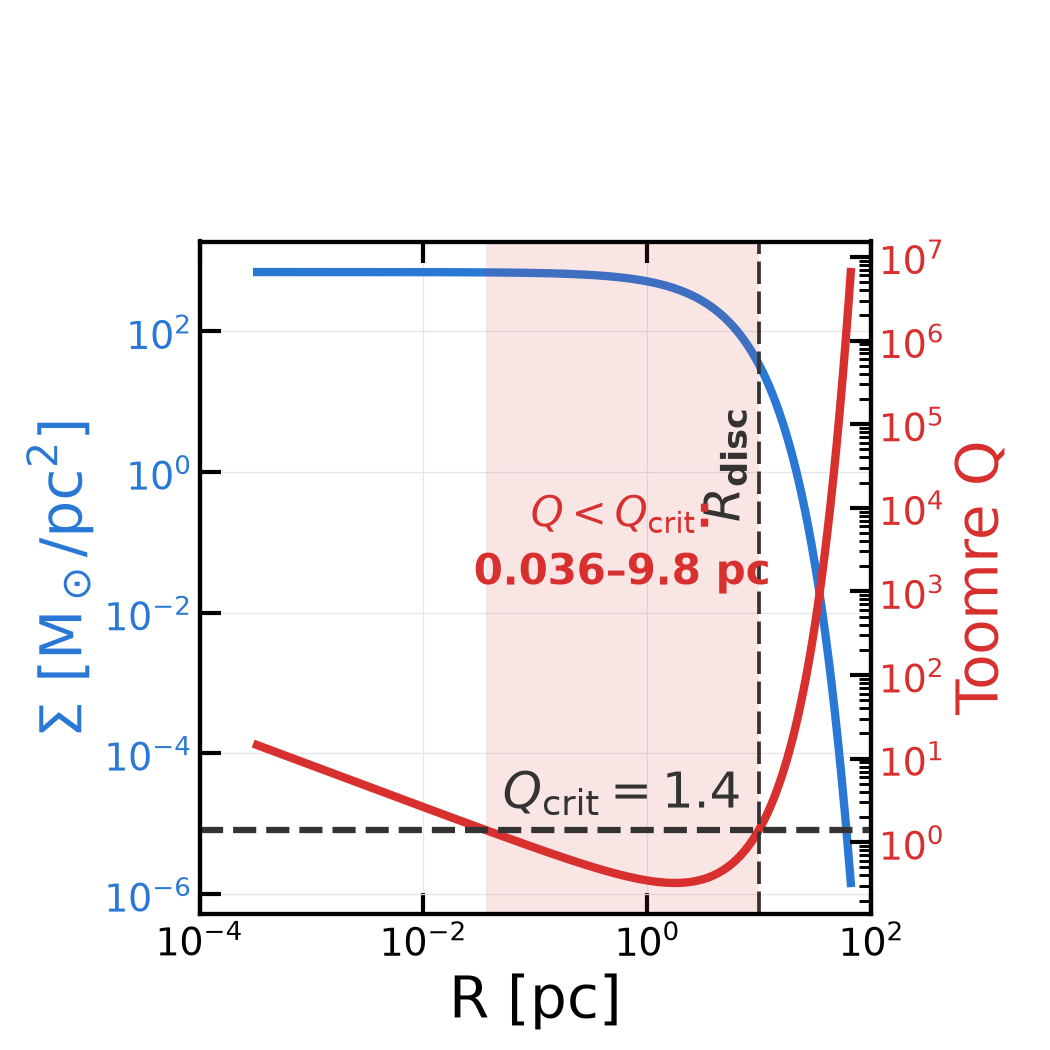}
        \label{fig:sigmaQ_ET_best}
    \end{subfigure}
    \begin{subfigure}[b]{0.325\textwidth}
        \centering
        \includegraphics[width=\linewidth]{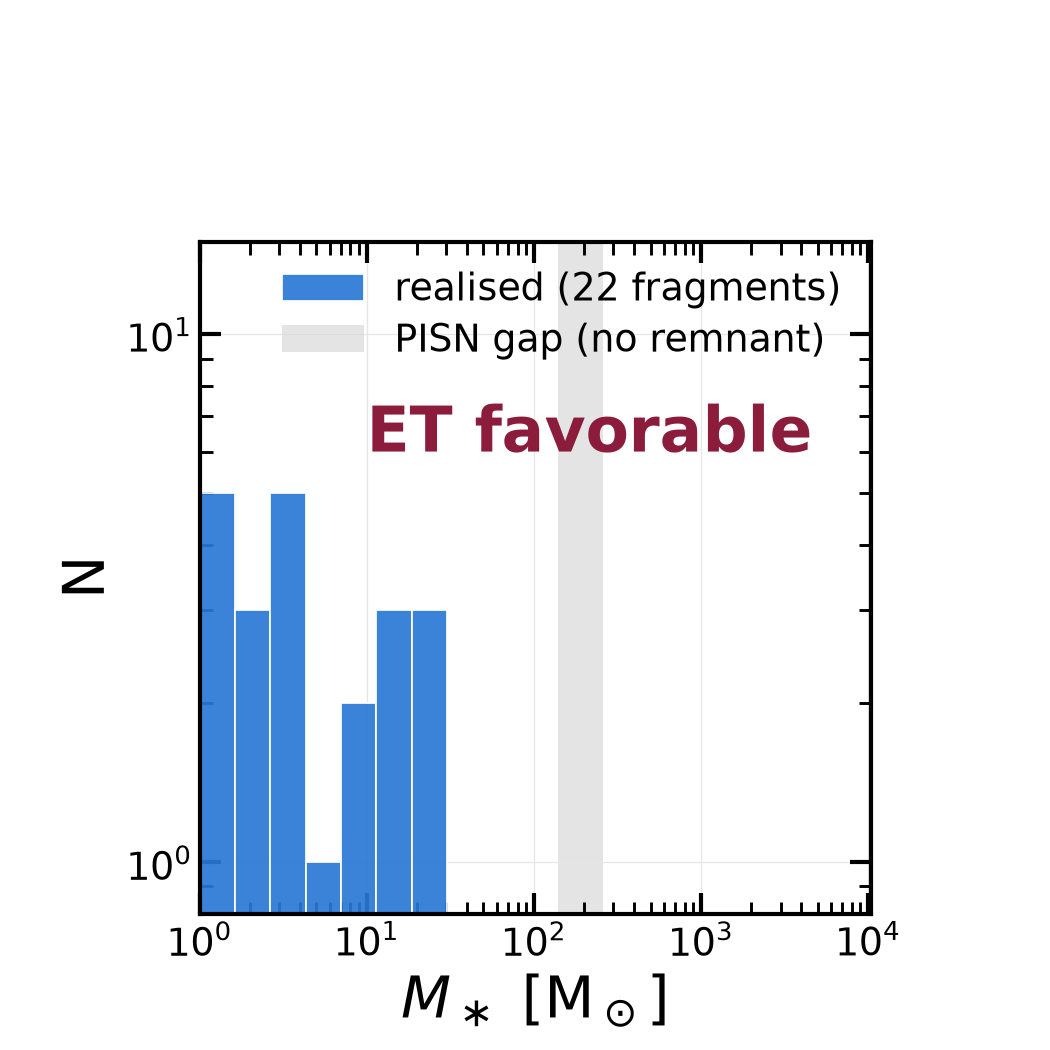}
        \label{fig:popIII_IMF_ET_best}
    \end{subfigure}
    \begin{subfigure}[b]{0.325\textwidth}
        \centering
        \includegraphics[width=\linewidth]{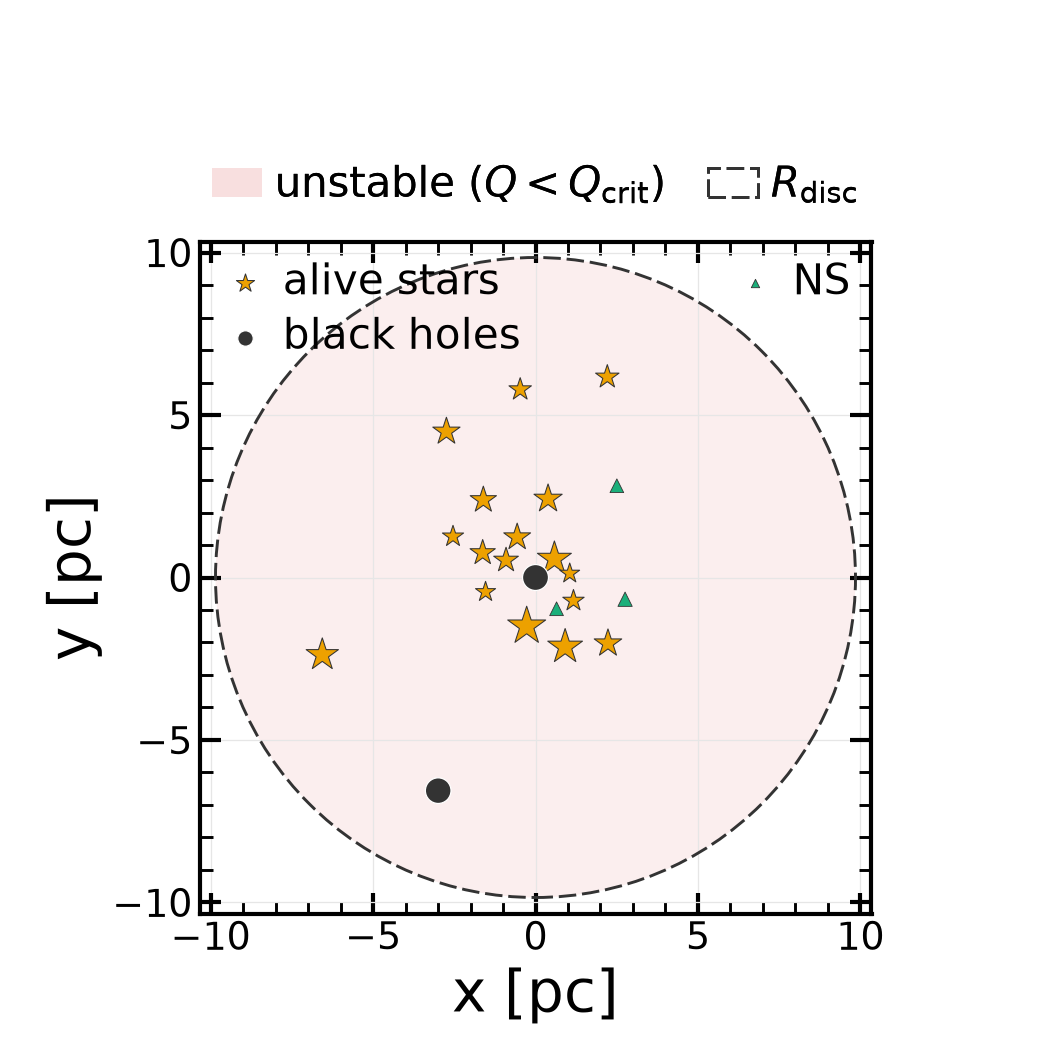}
        \label{fig:cluster_ET_best}
    \end{subfigure}
    \centering
    \begin{subfigure}[b]{0.325\textwidth}
        \centering
        \includegraphics[width=\linewidth]{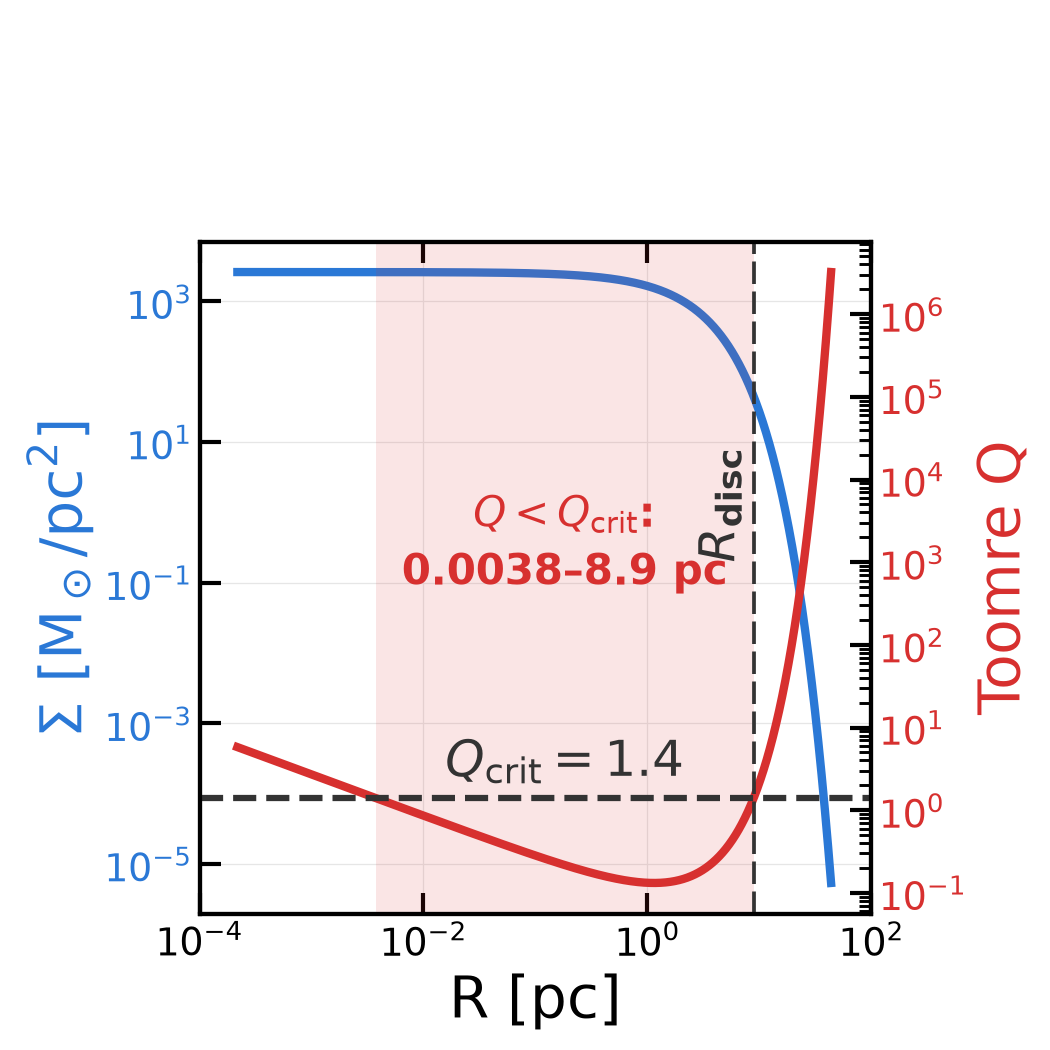}
        \caption{}
        \label{fig:sigmaQ_LISA_best}
    \end{subfigure}
    \hfill
    \begin{subfigure}[b]{0.325\textwidth}
        \centering
        \includegraphics[width=\linewidth]{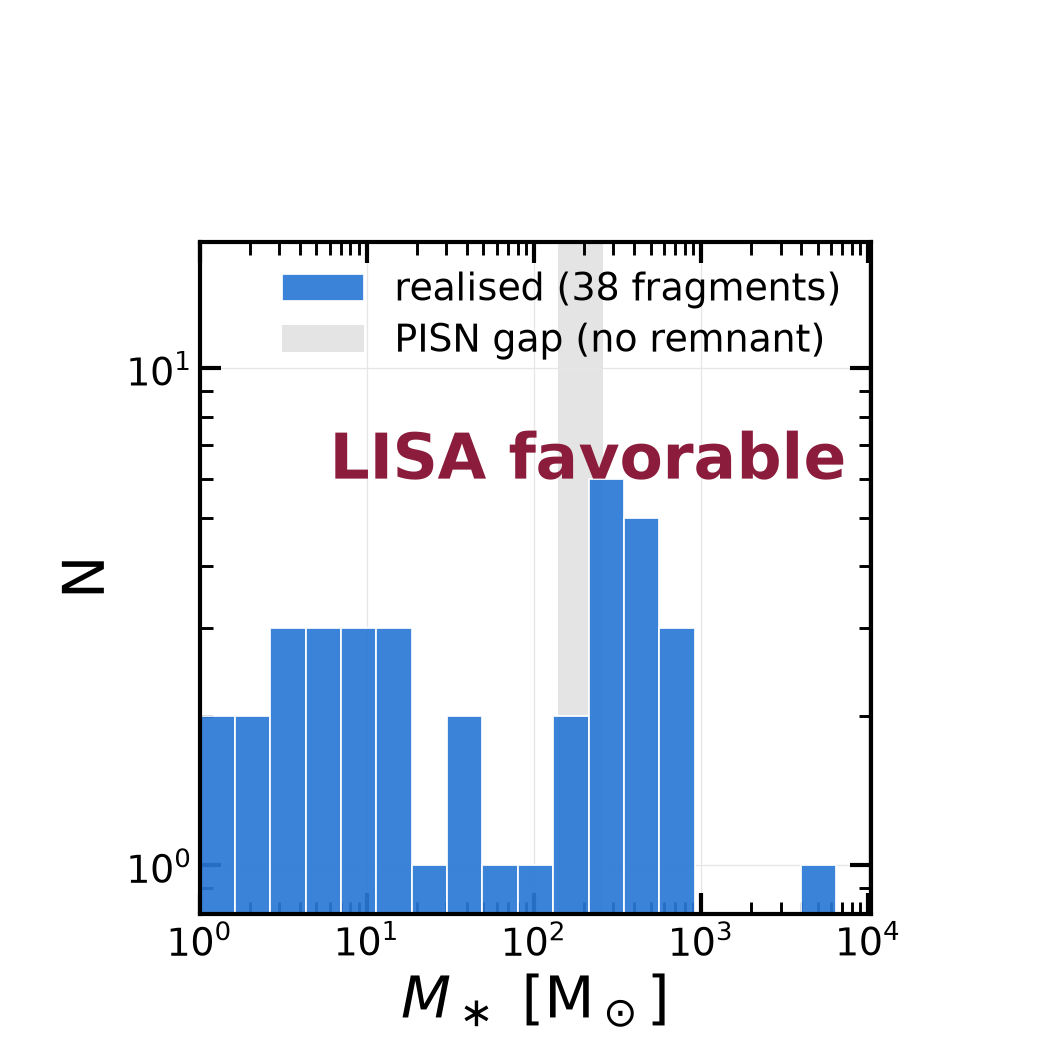}
        \caption{}
        \label{fig:popIII_IMF_LISA_best}
    \end{subfigure}
    \begin{subfigure}[b]{0.325\textwidth}
        \centering
        \includegraphics[width=\linewidth]{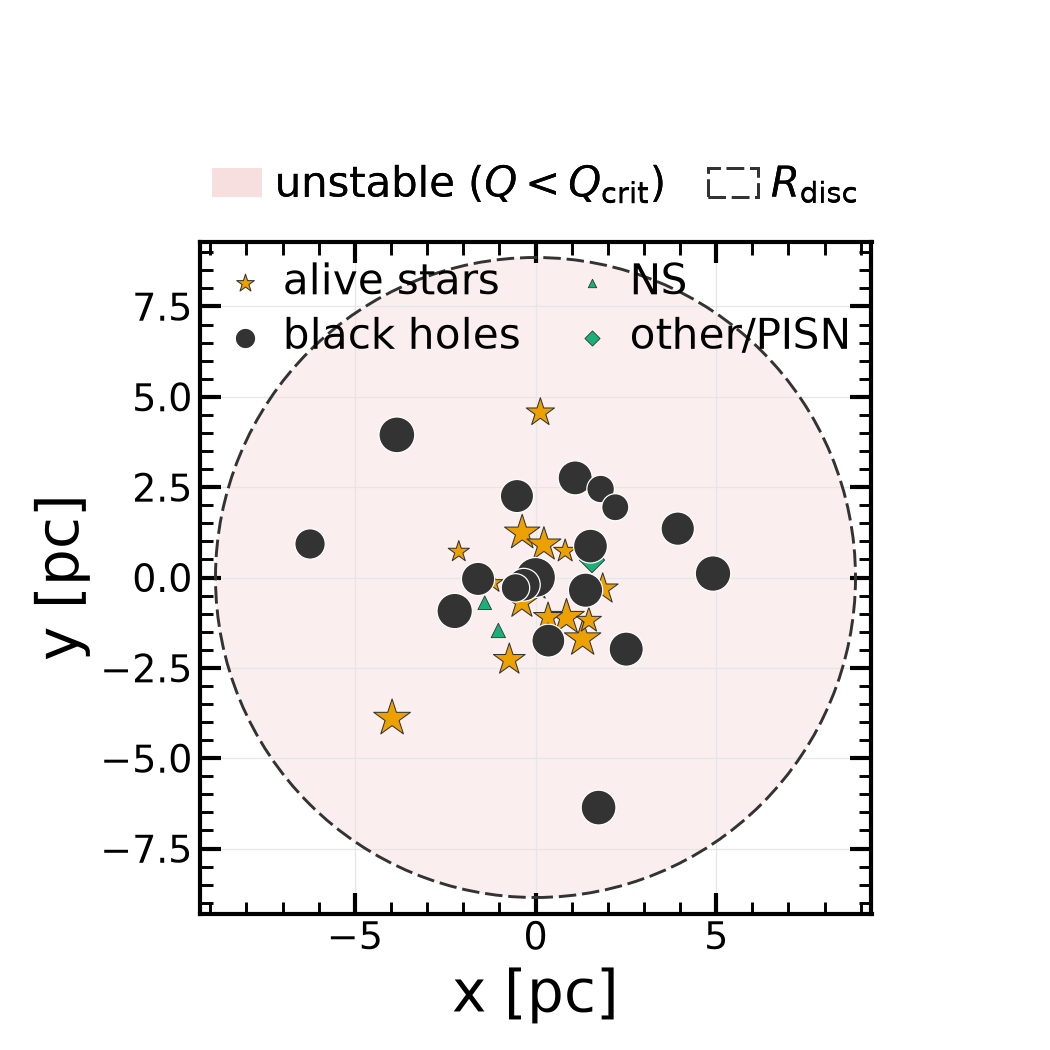}
        \caption{}
        \label{fig:cluster_LISA_best}
    \end{subfigure}
    \caption{Initial conditions of the Pop~III seeding model for parameter combinations yielding GW signals detectable by ET (top panels) and LISA (bottom panels). 
    \textbf{(a)}~Radial profiles of the gas surface density $\Sigma$ (blue) and the Toomre parameter $Q$ (red). The dashed line marks $Q_{\rm crit}=1.4$ and the shaded band the gravitationally unstable region ($Q<Q_{\rm crit}$), where the stars form.
    \textbf{(b)}~Distribution of the initial stellar masses of one realization, drawn from the top-heavy IMF $\mathrm{d}N/\mathrm{d}M\propto M^{-\alpha}$, truncated at $\epsilon_{\rm clump} M_{\rm T}(R)$ (the Toomre mass at the formation radius), as the number of stars per logarithmic mass bin.
    \textbf{(c)}~Face-on spatial distribution of the stars and their remnants $10\,$Myr after the star formation. Stars still alive are shown as stars, BHs as circles. The symbol size increases with mass, and the shaded disk shows the unstable region of panel~(a).}
\end{figure*}

\begin{figure}[t] 
    \centering 
    \begin{subfigure}{\linewidth} \centering            \includegraphics[width=\linewidth]{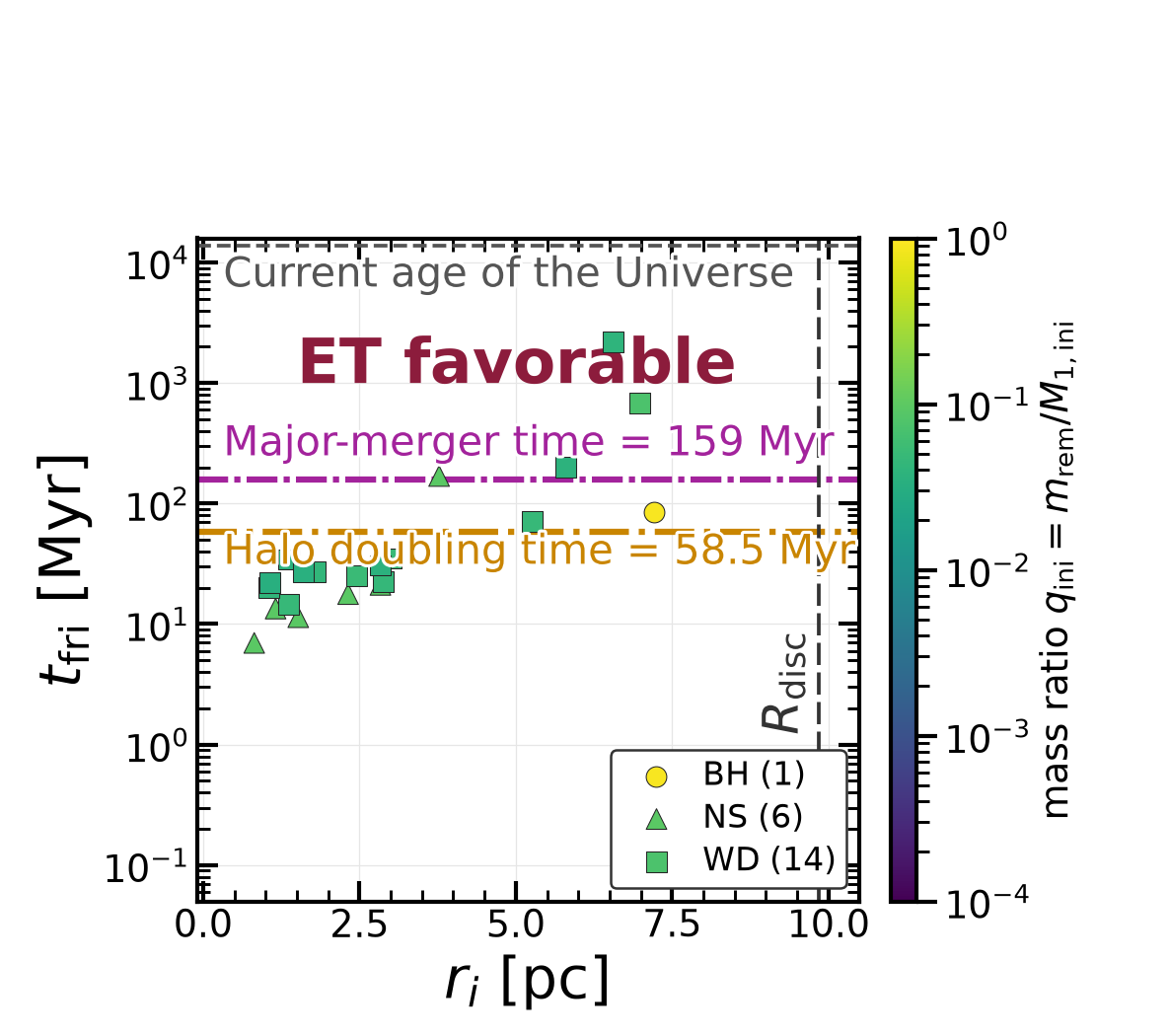} 
    \label{fig:dynamical_friction_ET_best} \end{subfigure} 
    \vspace{0.01em} 
    \begin{subfigure}{\linewidth} 
    \centering 
    \includegraphics[width=\linewidth]{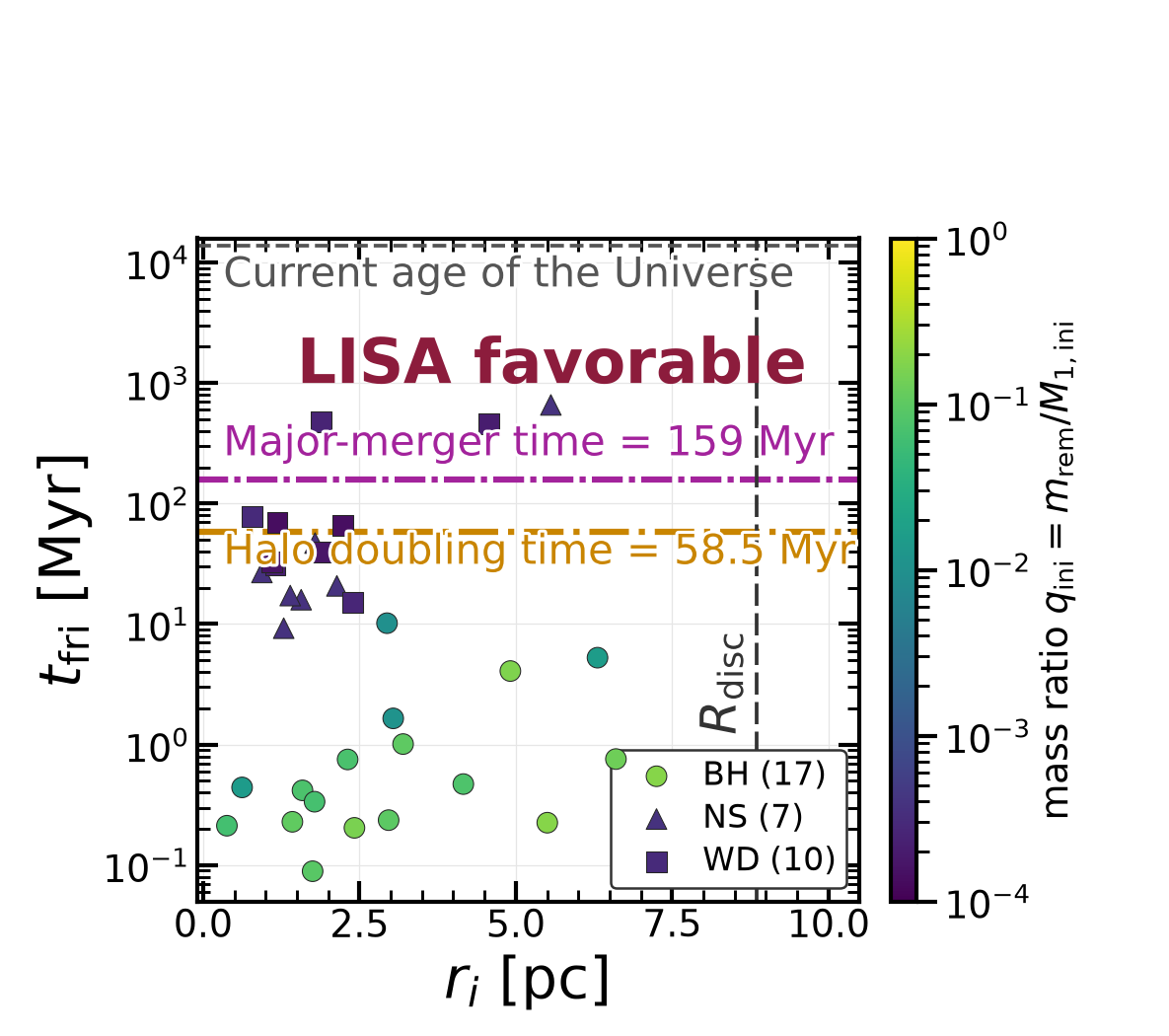} \label{fig:dynamical_friction_LISA_best} 
    \end{subfigure}
    \caption{Dynamical-friction time $t_{\rm fri}$ versus initial radius for the satellite remnants in the Pop~III model with $M_{\rm vir}=10^{6}\,\mathrm{M}_\odot$. Marker shapes denote the remnant type, while the color indicates the mass ratio $q_{\rm ini}=m_{\rm rem}/M_{1,\rm ini}$ relative to the primary's initial remnant mass. The horizontal lines mark the halo-doubling time and the major-merger time from \citet{Fakhouri2010}. Top panel: ET favorable parameters. Bottom panel: LISA favorable parameters.} 
    \label{fig:dynamical_friction}
\end{figure}

\begin{figure*}
    \centering
    \includegraphics[width=0.8\linewidth]{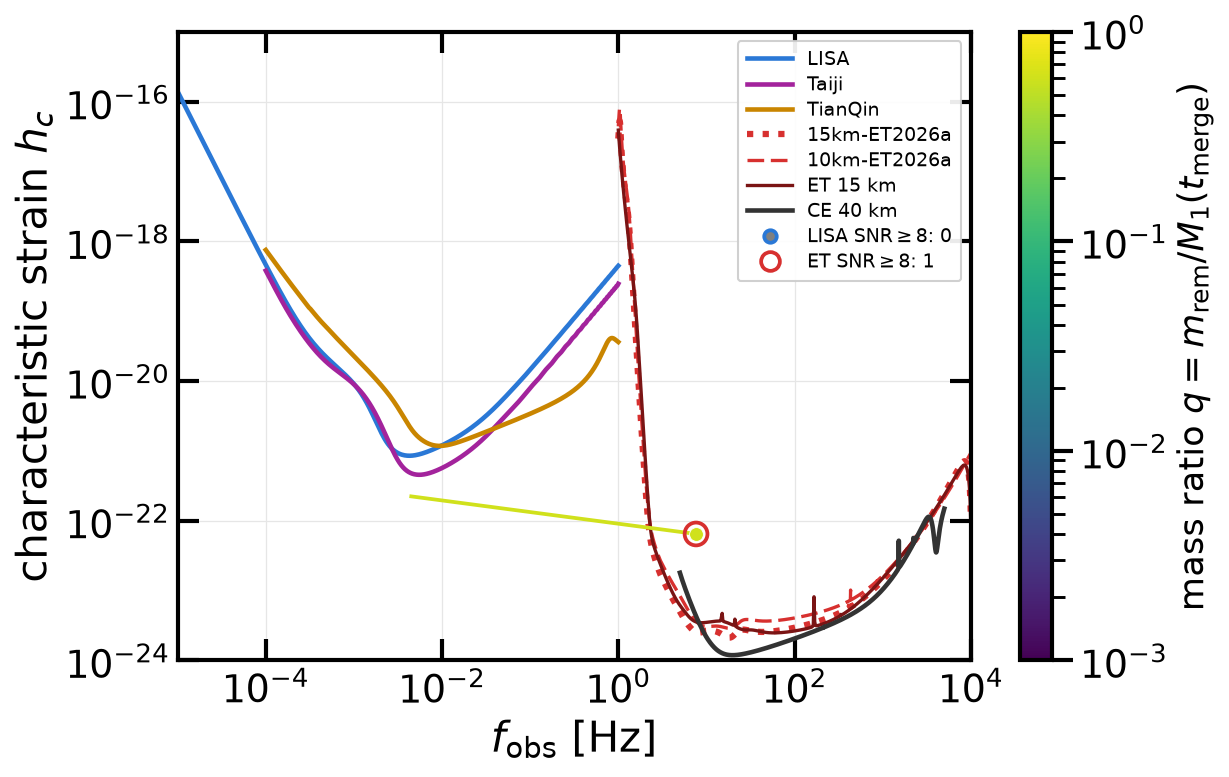}
    \caption{Characteristic strain $h_{\rm c}$ of the BH--BH mergers for the favorable ET parameters, compared with the detector sensitivities. The line segment shows the inspiral of one satellite remnant into the central BH, drawn from the frequency reached four years before it reaches $r_{\rm ISCO}$ to the observed ISCO frequency (circle), at the redshift of its own merger. The color gives the mass ratio $q=m_{\rm rem}/M_1(t_{\rm merge})$, with $M_1(t_{\rm merge})$ the primary mass just before the merger. The red ring marks an event with SNR$\,\geq8$ in ET. The solid dark-red line shows the ET sensitivity curve \citep{Branchesi2023} with its dashed and dotted red variants for the ET-2026a $10\,$km and $15\,$km configurations \citep{Dupletsa2026}. The LISA \citep{Robson2019}, Taiji \citep{Luo2021}, TianQin \citep{Mei2021}, and CE \citep{Hall2022} sensitivity curves are depicted by the blue, purple, amber, and black line, respectively.}
    \label{fig:GW_strain_ET_best}
\end{figure*}

We calculate the GW signals produced by the mergers of satellite BH remnants with the central primary. Each merger is approximated as a circular BH binary with primary mass $M_1$ and satellite mass $m_{\rm i}$. We evaluate its detectability with LISA, ET, and other GW detectors using the leading-order inspiral waveform. The source-frame chirp mass and GW frequency at the innermost stable circular orbit (ISCO) are, respectively,

\begin{equation}
\mathcal{M}_{\rm c}=
\frac{(M_1m)^{3/5}}{(M_1+m)^{1/5}},
\qquad
f_{\rm ISCO}=
\frac{c^3}{6^{3/2}\pi G (M_1+m)}\,.
\label{eq:chirp}
\end{equation}

We assume a non-spinning Schwarzschild \citep{Bardeen1972} approximation for the ISCO frequency. Cosmological redshift is included through the observed chirp mass $\mathcal{M}_{\rm c,z}=(1+z)\mathcal{M}_{\rm c}$, the observed frequency $f_{\rm obs}=f/(1+z)$, and the luminosity distance,

\begin{equation}
D_{\rm L}(z)=(1+z)\frac{c}{H_0}
\int_0^z
\frac{{\rm d}z'}
{\sqrt{\Omega_{\rm m}(1+z')^3+\Omega_\Lambda}}.
\label{eq:dl}
\end{equation}

For each merger, we use the mass of the central BH immediately before coalescence, accounting for its growth through gas accretion and previous BH mergers. We also determine the merger redshift from the cosmic time corresponding to each event. 

In the stationary-phase approximation, the leading-order frequency-domain inspiral amplitude is

\begin{equation}
|\tilde h(f)|=
\sqrt{\frac{5}{24}}\,\pi^{-2/3}
\frac{(G\mathcal{M}_{\rm c,z})^{5/6}}
{c^{3/2}D_{\rm L}}f^{-7/6},
\label{eq:htilde}
\end{equation}

\noindent where $f$ is the observed GW frequency. The corresponding characteristic strain is $h_{\rm c}(f)=2f|\tilde h(f)|\propto f^{-1/6}$.

We calculate the matched-filter signal-to-noice ratio (SNR) for each detector as

\begin{equation}
{\rm SNR}^2=
\mathcal{A}\int_{f_{\rm start}}^{f_{\rm ISCO,obs}}
\frac{|\tilde h(f)|^2}{S_n(f)}\,{\rm d}f,
\label{eq:snr}
\end{equation}

\noindent where $f_{\rm ISCO,obs}=f_{\rm ISCO}/(1+z)$ and $\mathcal{A}$ converts the optimal (face-on, overhead) amplitude of Equation~\eqref{eq:htilde} into an SNR averaged over sky position, polarization, and inclination. The LISA, TianQin, and Taiji sensitivity curves are already averaged over sky position and polarization, so only the inclination average remains, giving $\mathcal{A}=16/5$ \citep{Robson2019}. The ET and CE noise curves refer to a single L-shaped interferometer with optimal orientation; averaging over all angles reduces the SNR by a factor $2/5$ (i.e.\ $\mathcal{A}=4\times(2/5)^2$) for CE, and by $\sqrt{2}\times2/5\simeq0.57$ for the ET 2L configuration of two 15\,km L-shaped detectors (i.e.\ $\mathcal{A}=8\times(2/5)^2$).  For space-based detectors, $f_{\rm start}$ is determined by $t(f_{\rm start})-t(f_{\rm ISCO,obs})=T_{\rm obs}$, with $T_{\rm obs}=4$ yr, using the leading-order inspiral relation,

\begin{equation}
t(f)=\frac{5}{256}
\left(\frac{G\mathcal{M}_{\rm c,z}}{c^3}\right)^{-5/3}
(\pi f)^{-8/3}.
\label{eq:tgw}
\end{equation}

For ET and CE, it is the low-frequency cutoff of the detector,
at $1$ and $5$\,Hz, respectively.

We assess GW detectability using the sensitivity curves of LISA \citep{Robson2019}, including instrumental and Galactic confusion noise, TianQin \citep{Mei2021}, Taiji \citep{Ruan2020}, ET \citep{Dupletsa2026}, combining its low- and high-frequency interferometers, and the 40 km CE configuration \citep{Hall2022}. The SNR is integrated over the final $T_{\rm obs}$ years of inspiral for space-based detectors and over the accessible inspiral band for ground-based detectors, terminating at the observed ISCO frequency. We set the SNR to zero when this frequency falls below the detector's lower cutoff and adopt ${\rm SNR}\geq8$ as the detection threshold. We consider only circular inspirals, neglecting BH spins, eccentricity, merger, and ringdown. Including these effects could modify the predicted SNRs, particularly for high-redshift binaries whose signals terminate near the low-frequency sensitivity limit of ground-based detectors \citep{Flanagan1998,Ajith2008,Khan2016}.

\section{Results}\label{sec:result}

Given the uncertainties and the differing sensitivities of space-based (LISA, TianQin, and Taiji) and ground-based (ET and CE) GW detectors to chirp mass, mass ratio, and redshift, we perform a proof-of-concept exploration of our parameter space. 

Specifically, we identify favourable parameter configurations that yield GW events detectable by ET and LISA (${\rm SNR}\geq8$), hereafter referred to as ET-detectable events (EDEs) and LISA-detectable events (LDEs), respectively. The corresponding parameters are listed in the second and third columns of Table~\ref{tab:best_params}. Since the model is stochastic, with varying IMF sampling and fragment positions, we perform 1000 independent realisations for each configuration. For illustration, we select the realisation producing the most LDEs (11) and one of the two realisations yielding an EDE, representing optimistic rather than typical outcomes.
We present the ET case first, followed by LISA. These examples
illustrate detectability in different GW frequency bands rather
than statistically representative predictions for the Pop~III
halo population, which we defer to future work.


\subsection{Detection with the Einstein Telescope}

The top panel of Figure~\ref{fig:sigmaQ_LISA_best} shows the radial profiles of the disk surface density (blue) and Toomre parameter (red) for favorable ET parameters in Table \ref{tab:best_params}, with the dashed line marking the critical value $Q_{\rm crit}$ below which the disk becomes gravitationally unstable. At small radii, the disk is stabilized by the high epicyclic frequency, whereas at large radii, the declining surface density suppresses gravitational instability. Fragmentation is therefore confined to an intermediate annulus where $Q<Q_{\rm crit}$.

\begin{figure*}
    \centering
    \includegraphics[width=0.8\linewidth]{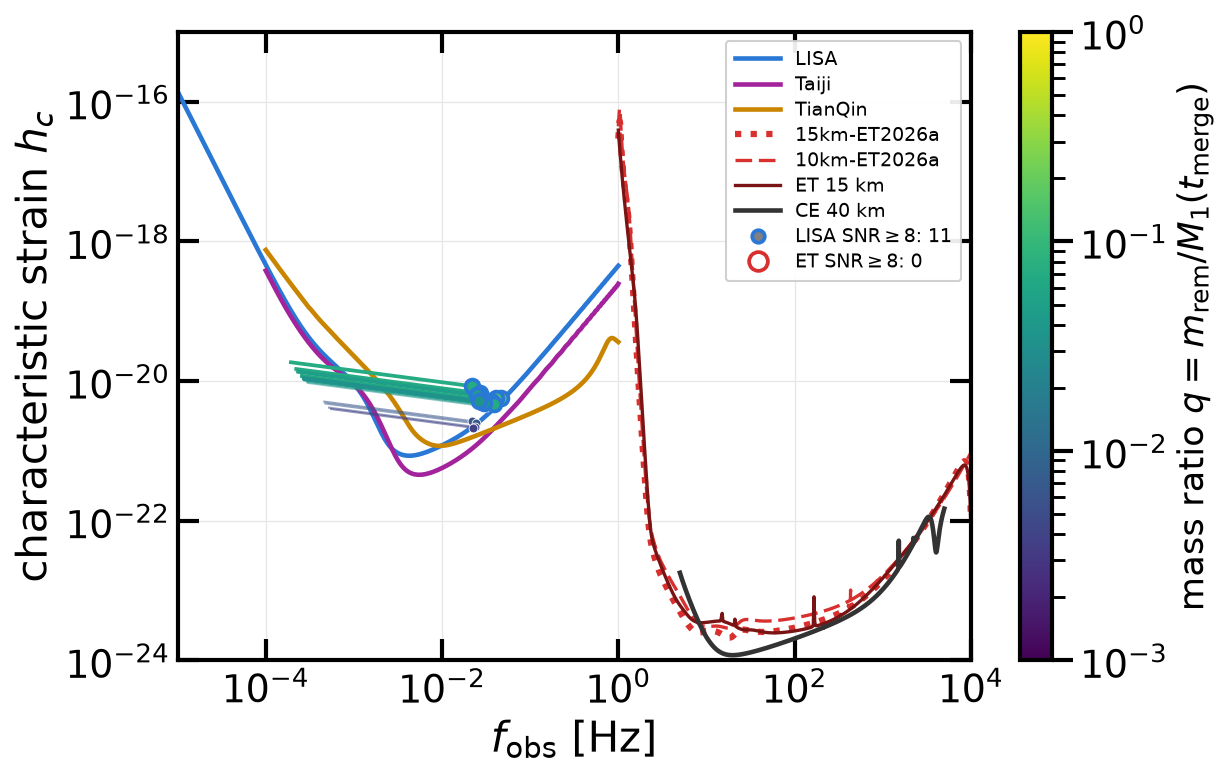}
    \caption{Characteristic strain $h_{\rm c}$ of the BH--BH mergers for the favorable LISA parameters, compared with the detector sensitivities. Each line segment is the inspiral of one satellite remnant into the central BH, drawn from the frequency reached four years before it reaches $r_{\rm ISCO}$ to the observed ISCO frequency (circle), at the redshift of its own merger. The color gives the mass ratio $q=m_{\rm rem}/M_1(t_{\rm merge})$, with $M_1(t_{\rm merge})$ the primary mass just before the merger. The blue ring marks an event with SNR$\,\geq8$ in LISA. Sensitivity curves are as in Figure~\ref{fig:GW_strain_ET_best}.}
    \label{fig:GW_strain_LISA_best}
\end{figure*}

The top panel of Figure~\ref{fig:popIII_IMF_LISA_best}  shows the resulting Pop~III stellar mass distribution. The relatively steep IMF slope of $\alpha=1.5$, together with the adopted clump efficiency $\epsilon_{\rm clump} = 0.1$, favors the formation of a larger number of comparatively low-mass Pop~III stars. The top panel of Figure~\ref{fig:cluster_LISA_best} shows their spatial distribution at $t_{\rm obs}=10$~Myr after the initial redshift of $z=20$. By this time, a few Pop~III stars have already evolved off the main sequence, while the remaining stars are still on the main sequence due to their lower masses and correspondingly longer lifetimes. 

Figure~\ref{fig:dynamical_friction} shows the dynamical-friction sinking time-scale, $t_{\rm fri}$, as a function of the initial orbital radius, $r_i$, for the stellar remnants. Circles, triangles, and squares represent BHs, NSs, and WDs, respectively, with the colors indicating their initial mass ratios, $q_{\rm ini}=m_{\rm rem}/M_{1,\rm ini}$. The vertical dashed line marks the disk radius. The orange and magenta dash-dotted horizontal lines indicate the characteristic halo mass-doubling and major-merger time-scales, respectively, estimated using the extrapolations from the results of \citet{Fakhouri2010}\footnote{This should be treated as an order-of-magnitude estimate, as the fitting relations of \citet{Fakhouri2010} were calibrated for halo masses of $10^{10}$--$10^{15}{\rm M_\odot}$ and redshifts $z\lesssim15$, whereas we extrapolate these relations to $M_{\rm vir}=10^6 {\rm M_\odot}$ at $z=20$.}, with major mergers defined by a progenitor mass ratio $\xi\geq0.3$. In the case of EDEs in the top panel of Figure~\ref{fig:dynamical_friction}, the BH remnant has sinking times significantly shorter than both halo evolution time-scales, suggesting that it can reach the central region before substantial halo growth or a major merger. In contrast, some lower-mass remnants, particularly WDs, exhibit considerably longer sinking times.

The halo mass-doubling time-scale from \citet{Fakhouri2010} is $\sim2$--$3$ times longer than that from \citet{Yung2024} for the masses and redshifts considered. However, large-scale halo accretion does not immediately affect the central disk. For a $10^6\,{\rm M_\odot}$ halo at $z=20$
($R_{\rm vir}\simeq150$ pc), material accreted from $\sim2R_{\rm vir}$ with an enclosed mass of $\sim2M_{\rm vir}$ has a free-fall time of $\sim60$ Myr. Allowing two free-fall times for infall and settling gives $\sim120$ Myr, suggesting that the central disk may remain approximately isolated beyond the halo mass-doubling time.
As we do not model the subsequent halo accretion and merger history, we adopt the major-merger time-scale as a practical upper limit and consider only inspirals completed within it.

Figure~\ref{fig:GW_strain_ET_best} shows the characteristic GW strain, $h_{\rm c}$, as a function of the observed frequency, $f_{\rm obs}$, for the BH binaries, together with the sensitivity curves of LISA, Taiji, TianQin, ET, and CE. Individual signals are colored according to the mass ratio, $q=m_{\rm rem}/M_1$, evaluated at the time of merger 
whereas the highlighted circles identify binaries with ${\rm SNR}\geq8$.

In the ET-detectable case, one binary reaches the detection threshold, whereas none are detectable by LISA. This system has a mass ratio at the time of merger 
close to unity and produces a signal extending into the ET frequency band at $f_{\rm obs}\sim8$~Hz. Although its inspiral signal also extends through the frequency range of space-based detectors, the characteristic strain remains below their sensitivity curves, resulting in an insufficient SNR for detection.
   
\subsection{Detection with the Laser Interferometer Space Antenna}

The bottom panel of Figure~\ref{fig:sigmaQ_LISA_best} shows the radial profiles of the gas surface density and Toomre parameter, whereas the bottom panel of Figure~\ref{fig:popIII_IMF_LISA_best} presents the resulting Pop~III stellar mass distribution for the favorable LISA parameters from Table \ref{tab:best_params}. The relatively small number of fragments (35) leads to significant stochastic variations, so the sampled mass distribution does not necessarily reproduce the underlying power-law IMF. Comparing this with the ET case (top panel of Figure~\ref{fig:popIII_IMF_LISA_best}), we find that mergers detectable by LISA are obtained with a more top-heavy IMF ($\alpha=0.8$), whereas the ET-detectable case corresponds to a steeper IMF ($\alpha=1.5$). This difference arises primarily from the distinct sensitivities of the two detectors to BH masses and mass ratios. LISA operates at lower GW frequencies, making it possible to detect mergers involving more massive primary BHs, which can form from a more top-heavy IMF and subsequent gas accretion. In contrast, ET is sensitive to higher GW frequencies, allowing it to detect mergers involving lower-mass BHs, particularly those with more comparable component masses.

The top and bottom panels of Figures~\ref{fig:cluster_LISA_best} show the spatial distributions at $t_{\rm obs}=10$~Myr after $z=20$. In the LISA case, the more massive Pop~III stars have shorter lifetimes, with most already evolving into compact remnants. In contrast, the ET case is dominated by lower-mass stars, with only few having left the main sequence. The initial mass of the central BH is also significantly more in the LISA case ($M_1\simeq3512\,{\rm M_\odot}$) than in ET ($M_1\simeq15.7\,{\rm M_\odot}$) 

\begin{table}
\centering
\small
\setlength{\tabcolsep}{3pt}
\caption{Events with orientation-averaged SNR$\,\geq8$ for the favorable
ET model  (top) and the favorable LISA model (bottom). $M_1$ is the primary mass just
before the merger and $q=m_{\rm rem}/M_{1}(t_{\rm merger})$ is the mass ratio to it, $z$ is the merger redshift, and the following columns show SNR values in each GW detector. ET uses the ET2026a 15\,km 2L
configuration; LISA, TianQin and Taiji SNRs are sky-, polarization- and
inclination-averaged.}
\label{tab:events}
\begin{tabular}{lccccccc}
\hline
$M_1\,[\mathrm{M}_\odot]$ & $q$ & $z_{\rm merge}$ &
LISA & ET & CE & TianQin & Taiji \\
\hline
\multicolumn{8}{c}{\textbf{EDEs}} \\
\hline
22 & 0.609 & 15.00 & 0.20 & 8.43 & 2.35 & 0.19 & 0.41 \\
\hline
\multicolumn{8}{c}{\textbf{LDEs}} \\
\hline
9072 & 0.069 & 19.55 & 13.70 & 0 & 0 & 8.55 & 23.46 \\
7365 & 0.066 & 19.79 & 11.29 & 0 & 0 & 7.19 & 19.35 \\
3807 & 0.185 & 19.84 & 10.73 & 0 & 0 & 7.10 & 18.43 \\
4513 & 0.132 & 19.84 & 10.54 & 0 & 0 & 6.93 & 18.09 \\
8167 & 0.047 & 19.77 & 10.40 & 0 & 0 & 6.57 & 17.82 \\
5467 & 0.073 & 19.83 & 9.28 & 0 & 0 & 6.05 & 15.94 \\
5872 & 0.063 & 19.83 & 9.13 & 0 & 0 & 5.93 & 15.67 \\
7054 & 0.041 & 19.80 & 8.63 & 0 & 0 & 5.54 & 14.81 \\
7884 & 0.034 & 19.78 & 8.61 & 0 & 0 & 5.47 & 14.76 \\
5115 & 0.068 & 19.84 & 8.45 & 0 & 0 & 5.54 & 14.51 \\
6774 & 0.041 & 19.80 & 8.30 & 0 & 0 & 5.34 & 14.24 \\
\hline
\end{tabular}
\end{table}

The bottom panel of Figure~\ref{fig:dynamical_friction} shows the dynamical friction time-scales of the stellar remnants as a function of their initial orbital radii, with colors indicating their mass ratios relative to the initial mass of central BH. In this realization, all 19 BH satellites have sinking times of $\lesssim30$~Myr, shorter than both the halo mass-doubling time-scale ($\sim58.5$~Myr) and the major-merger time-scale ($\sim159$~Myr). Thus, all 19 BH remnants can merge with the central primary within the adopted isolated-halo evolution window. In contrast, the lower-mass NSs and WDs generally have longer sinking times, with some exceeding the major-merger time-scale.

Figure~\ref{fig:GW_strain_LISA_best} shows the characteristic GW strains of these 19 BH mergers compared with the sensitivity curves of future GW detectors. Of these, 11 events satisfy ${\rm SNR}\geq8$ for LISA, whereas none are detectable by ET. The detectable signals are concentrated around observed frequencies of $\sim10^{-2}$~Hz and terminate below the ET frequency band. In contrast, the ET-favorable case (Figure~\ref{fig:GW_strain_ET_best}) produces one ET-detectable merger and no LISA detections. These results highlight the complementary sensitivities of space-based and ground-based GW observatories to different BH mass and mass-ratio regimes.

Table~\ref{tab:events} presents the SNRs of the one ET-detectable and 11 LISA-detectable mergers across five GW detectors for single best IMF realization. The ET-favorable event, involving nearly equal-mass BHs at $z_{\rm merge}=15$, is detectable only by ET. 
In contrast, all 11 LISA-detectable mergers at $z_{\rm merge}\sim19$--$20$ also satisfy ${\rm SNR}\geq8$ for Taiji, while only one reaches this threshold for TianQin. None of these events is detectable by ET or CE, further highlighting the complementary sensitivity of space-based and ground-based GW observatories to different BH merger populations.

\section{Discussion and conclusion}\label{sec:discussion}

In this work, we have developed a semi-analytical model that follows the evolution of primordial gas disks, Pop~III star formation, and the subsequent inspiral and merger of their BH remnants to predict the resulting GW signals. As a proof of concept, we selected favorable parameter configurations from our model grid specifically to produce events potentially detectable by ET or LISA. These examples clearly show that Pop~III remnant BH mergers can, under suitable astrophysical conditions, enter the sensitivity bands of both detectors; they should not, however, be interpreted as representative predictions for the overall population or its detection rate. ET and LISA probe different regions of the BH mass and mass-ratio parameter space, highlighting their complementary potential for detecting Pop~III remnant mergers at high redshift.

In our model, ET-detectable events typically involve primary BHs with masses of $M_1\lesssim$ a few tens of ${\rm M_\odot}$, allowing their inspiral signals to reach the ET frequency band above $\sim1$ Hz. However, these systems produce insufficient signal strength in the LISA band to satisfy the detection threshold of ${\rm SNR}\geq8$. In contrast, LISA-detectable mergers involve more massive primary BHs, with $M_1\sim10^3$--$10^4{\rm M_\odot}$. For the most massive systems, the observed ISCO frequencies can be as low as $f_{\rm ISCO,obs}\sim0.01$--$0.02$ Hz, below the lower-frequency limit of ET.

The Eddington ratio $f_{\rm Edd}$ also plays a role in determining detectability, as it controls the growth of the central BH through gas accretion. Higher accretion rates ($f_{\rm Edd}\sim1$) allow the primary to grow to larger masses, shifting the GW signal towards lower frequencies and favoring detection by LISA. In contrast, lower accretion rates ($f_{\rm Edd}\sim0.01$) keep the primary BH relatively less massive, making its mergers more favorable for a detection by ET. These differences explain why the two detectors favor distinct combinations of IMF slope and accretion rate in our model. A systematic investigation of the relative importance of these parameters and their impact on the overall number of detectable events is left for future work.

Numerical simulations suggest that a significant fraction of Pop~III stars may form in binaries or multiple systems \citep{Stacy2013, Stacy2016, Susa2014}. Although we do not include binary evolution within the satellite population, such systems could provide an additional channel for hierarchical BH mergers. In particular, close binaries containing comparable-mass BHs may first merge and produce GW signals detectable by ET. If the resulting remnant remains bound to the host halo, it could subsequently inspiral towards the more massive central BH and undergo another merger, potentially detectable by LISA. Such a scenario could produce successive GW events in different frequency bands from the same evolutionary system. However, its occurrence depends on the efficiency of binary hardening, the retention of merger remnants against GW recoil, and their subsequent migration towards the central BH. Including these effects in future work would allow us to investigate whether Pop~III binaries can provide an additional connection between the merger populations accessible to ET and LISA.

Moreover, these results motivate a broader exploration of the astrophysical parameter space governing Pop~III star formation and BH remnant evolution. In particular, the relative number of detectable mergers across GW observatories operating in different frequency bands could help constrain the properties of primordial disks, the Pop~III initial mass function, and the subsequent growth of BH remnants. However, our present model predicts the number of detectable mergers per individual halo rather than cosmological detection rates. Estimating the latter requires accounting for the abundance and redshift evolution of host mini-haloes, their hierarchical assembly and merger histories, and the selection effects of different GW detectors. Incorporating these effects in future work will allow us to assess the cosmological event rates and the potential of multi-band GW observations to constrain the formation and evolution of the first generation of stars and their BH remnants.

\begin{acknowledgments}
{LM thanks Ralf Klessen for useful discussions. PRC acknowledges support from the Swiss National Science Foundation
under the Sinergia Grant CRSII5\_213497 (GW-Learn).}
\end{acknowledgments}
\software{We acknowledge usage of the Python programming language \citep{python_van1995}, Astropy \citep{astropy2018}, Matplotlib \citep{matplotlib_hunter2007}, and Numpy \citep{numpy_harris2020}.}



\bibliography{ref}{}

@article{astropy2018,
  title = {The astropy project: Building an open-science project and status of the v2. 0 core package},
  author = {Astropy Collaboration and Price-Whelan, a AM and Sip{\H{o}}cz, BM and G{\"u}nther, HM and Lim, PL and Crawford, SMea and Conseil, S and Shupe, DL and Craig, MW and Dencheva, N and others},
  journal = {The Astronomical Journal},
  volume = {156},
  number = {3},
  pages = {123},
  year = {2018},
  publisher = {The American Astronomical Society},
}

@article{Derdzinski2023,
  author = {{Derdzinski}, Andrea and {Mayer}, Lucio},
  title = {{In situ extreme mass ratio inspirals via subparsec formation and migration of stars in thin, gravitationally unstable AGN discs}},
  journal = {\mnras},
  year = {2023},
  month = may,
  volume = {521},
  number = {3},
  pages = {4522-4543},
  doi = {10.1093/mnras/stad749},
  archiveprefix = {arXiv},
  eprint = {2205.10382},
  primaryclass = {astro-ph.GA},
  adsurl = {https://ui.adsabs.harvard.edu/abs/2023MNRAS.521.4522D},
}

@article{Bullock2001,
  author = {{Bullock}, J.~S. and {Dekel}, A. and {Kolatt}, T.~S. and {Kravtsov}, A.~V. and {Klypin}, A.~A. and {Porciani}, C. and {Primack}, J.~R.},
  title = {{A Universal Angular Momentum Profile for Galactic Halos}},
  journal = {\apj},
  year = {2001},
  month = jul,
  volume = {555},
  number = {1},
  pages = {240-257},
  doi = {10.1086/321477},
  archiveprefix = {arXiv},
  eprint = {astro-ph/0011001},
  primaryclass = {astro-ph},
  adsurl = {https://ui.adsabs.harvard.edu/abs/2001ApJ...555..240B},
}

@article{Mo1998,
  author = {{Mo}, H.~J. and {Mao}, Shude and {White}, Simon D.~M.},
  title = {{The formation of galactic discs}},
  journal = {\mnras},
  year = {1998},
  month = apr,
  volume = {295},
  number = {2},
  pages = {319-336},
  doi = {10.1046/j.1365-8711.1998.01227.x},
  archiveprefix = {arXiv},
  eprint = {astro-ph/9707093},
  primaryclass = {astro-ph},
  adsurl = {https://ui.adsabs.harvard.edu/abs/1998MNRAS.295..319M},
}

@article{Schaerer2002,
  author = {{Schaerer}, D.},
  title = {{On the properties of massive Population III stars and metal-free stellar populations}},
  journal = {\aap},
  year = {2002},
  month = jan,
  volume = {382},
  pages = {28-42},
  doi = {10.1051/0004-6361:20011619},
  archiveprefix = {arXiv},
  eprint = {astro-ph/0110697},
  primaryclass = {astro-ph},
  adsurl = {https://ui.adsabs.harvard.edu/abs/2002A&A...382...28S},
}

@article{Heger2002,
  author = {{Heger}, A. and {Woosley}, S.~E.},
  title = {{The Nucleosynthetic Signature of Population III}},
  journal = {\apj},
  year = {2002},
  month = mar,
  volume = {567},
  number = {1},
  pages = {532-543},
  doi = {10.1086/338487},
  archiveprefix = {arXiv},
  eprint = {astro-ph/0107037},
  primaryclass = {astro-ph},
  adsurl = {https://ui.adsabs.harvard.edu/abs/2002ApJ...567..532H},
}

@article{Heger2003,
  author = {{Heger}, A. and {Fryer}, C.~L. and {Woosley}, S.~E. and {Langer}, N. and {Hartmann}, D.~H.},
  title = {{How Massive Single Stars End Their Life}},
  journal = {\apj},
  year = {2003},
  month = jul,
  volume = {591},
  number = {1},
  pages = {288-300},
  doi = {10.1086/375341},
  archiveprefix = {arXiv},
  eprint = {astro-ph/0212469},
  primaryclass = {astro-ph},
  adsurl = {https://ui.adsabs.harvard.edu/abs/2003ApJ...591..288H},
}

@article{Stacy2016,
  author = {{Stacy}, Athena and {Bromm}, Volker and {Lee}, Aaron T.},
  title = {{Building up the Population III initial mass function from cosmological initial conditions}},
  journal = {\mnras},
  year = {2016},
  month = oct,
  volume = {462},
  number = {2},
  pages = {1307-1328},
  doi = {10.1093/mnras/stw1728},
  archiveprefix = {arXiv},
  eprint = {1603.09475},
  primaryclass = {astro-ph.GA},
  adsurl = {https://ui.adsabs.harvard.edu/abs/2016MNRAS.462.1307S},
}

@article{Stacy2013,
  author = {{Stacy}, Athena and {Bromm}, Volker},
  title = {{Constraining the statistics of Population III binaries}},
  journal = {\mnras},
  year = {2013},
  month = aug,
  volume = {433},
  number = {2},
  pages = {1094-1107},
  doi = {10.1093/mnras/stt789},
  archiveprefix = {arXiv},
  eprint = {1211.1889},
  primaryclass = {astro-ph.CO},
  adsurl = {https://ui.adsabs.harvard.edu/abs/2013MNRAS.433.1094S},
}

@article{Susa2014,
  author = {{Susa}, Hajime and {Hasegawa}, Kenji and {Tominaga}, Nozomu},
  title = {{The Mass Spectrum of the First Stars}},
  journal = {\apj},
  year = {2014},
  month = sep,
  volume = {792},
  number = {1},
  eid = {32},
  pages = {32},
  doi = {10.1088/0004-637X/792/1/32},
  archiveprefix = {arXiv},
  eprint = {1407.1374},
  primaryclass = {astro-ph.GA},
  adsurl = {https://ui.adsabs.harvard.edu/abs/2014ApJ...792...32S},
}

@article{Hirano2014,
  author = {{Hirano}, Shingo and {Hosokawa}, Takashi and {Yoshida}, Naoki and {Umeda}, Hideyuki and {Omukai}, Kazuyuki and {Chiaki}, Gen and {Yorke}, Harold W.},
  title = {{One Hundred First Stars: Protostellar Evolution and the Final Masses}},
  journal = {\apj},
  year = {2014},
  month = feb,
  volume = {781},
  number = {2},
  eid = {60},
  pages = {60},
  doi = {10.1088/0004-637X/781/2/60},
  archiveprefix = {arXiv},
  eprint = {1308.4456},
  primaryclass = {astro-ph.CO},
  adsurl = {https://ui.adsabs.harvard.edu/abs/2014ApJ...781...60H},
}

@article{Clark2011,
  author = {{Clark}, Paul C. and {Glover}, Simon C.~O. and {Klessen}, Ralf S. and {Bromm}, Volker},
  title = {{Gravitational Fragmentation in Turbulent Primordial Gas and the Initial Mass Function of Population III Stars}},
  journal = {\apj},
  year = {2011},
  month = feb,
  volume = {727},
  number = {2},
  eid = {110},
  pages = {110},
  doi = {10.1088/0004-637X/727/2/110},
  archiveprefix = {arXiv},
  eprint = {1006.1508},
  primaryclass = {astro-ph.GA},
  adsurl = {https://ui.adsabs.harvard.edu/abs/2011ApJ...727..110C},
}

@article{Robson2019,
  author = {{Robson}, Travis and {Cornish}, Neil J. and {Liu}, Chang},
  title = {{The construction and use of LISA sensitivity curves}},
  journal = {Classical and Quantum Gravity},
  year = {2019},
  month = may,
  volume = {36},
  number = {10},
  eid = {105011},
  pages = {105011},
  doi = {10.1088/1361-6382/ab1101},
  archiveprefix = {arXiv},
  eprint = {1803.01944},
  primaryclass = {astro-ph.HE},
  adsurl = {https://ui.adsabs.harvard.edu/abs/2019CQGra..36j5011R},
}

@BOOK{Binney2008,
       author = {{Binney}, James and {Tremaine}, Scott},
        title = "{Galactic Dynamics: Second Edition}",
         year = 2008,
    publisher = "Princeton University Press, Princeton",
       adsurl = {https://ui.adsabs.harvard.edu/abs/2008gady.book.....B}
}

@article{Sasaki2014,
  author = {{Sasaki}, Mei and {Clark}, Paul C. and {Springel}, Volker and {Klessen}, Ralf S. and {Glover}, Simon C.~O.},
  title = {{Statistical properties of dark matter mini-haloes at z {\ensuremath{\geq}} 15}},
  journal = {\mnras},
  year = {2014},
  month = aug,
  volume = {442},
  number = {3},
  pages = {1942-1955},
  doi = {10.1093/mnras/stu985},
  archiveprefix = {arXiv},
  eprint = {1405.4018},
  primaryclass = {astro-ph.CO},
  adsurl = {https://ui.adsabs.harvard.edu/abs/2014MNRAS.442.1942S},
}

@article{Wu2026,
  author = {{Wu}, Jiaxi and {Most}, Elias R. and {Vu}, Nils L. and {Deppe}, Nils and {Kidder}, Lawrence E. and {Nelli}, Kyle C. and {Throwe}, William},
  title = {{Eccentricity as a Signature of Hierarchical Subsolar-mass Mergers in Collapsar Disks}},
  journal = {\apjl},
  year = {2026},
  month = jun,
  volume = {1004},
  number = {1},
  eid = {L19},
  pages = {L19},
  doi = {10.3847/2041-8213/ae743a},
  archiveprefix = {arXiv},
  eprint = {2604.26912},
  primaryclass = {astro-ph.HE},
  adsurl = {https://ui.adsabs.harvard.edu/abs/2026ApJ..1004L..19W},
}

@article{Greif2012,
  author = {{Greif}, Thomas H. and {Bromm}, Volker and {Clark}, Paul C. and {Glover}, Simon C.~O. and {Smith}, Rowan J. and {Klessen}, Ralf S. and {Yoshida}, Naoki and {Springel}, Volker},
  title = {{Formation and evolution of primordial protostellar systems}},
  journal = {\mnras},
  year = {2012},
  month = jul,
  volume = {424},
  number = {1},
  pages = {399-415},
  doi = {10.1111/j.1365-2966.2012.21212.x},
  adsurl = {https://ui.adsabs.harvard.edu/abs/2012MNRAS.424..399G},
}

@article{Hirano2017,
  author = {{Hirano}, Shingo and {Bromm}, Volker},
  title = {{Formation and survival of Population III stellar systems}},
  journal = {\mnras},
  year = {2017},
  month = sep,
  volume = {470},
  number = {1},
  pages = {898-914},
  doi = {10.1093/mnras/stx1220},
  archiveprefix = {arXiv},
  eprint = {1612.06387},
  primaryclass = {astro-ph.GA},
  adsurl = {https://ui.adsabs.harvard.edu/abs/2017MNRAS.470..898H},
}

@article{Burkert2016,
  author = {{Burkert}, A. and {F{\"o}rster Schreiber}, N.~M. and {Genzel}, R. and {Lang}, P. and {Tacconi}, L.~J. and {Wisnioski}, E. and {Wuyts}, S. and {Bandara}, K. and {Beifiori}, A. and {Bender}, R. and {Brammer}, G. and {Chan}, J. and {Davies}, R. and {Dekel}, A. and {Fabricius}, M. and {Fossati}, M. and {Kulkarni}, S. and {Lutz}, D. and {Mendel}, J.~T. and {Momcheva}, I. and {Nelson}, E.~J. and {Naab}, T. and {Renzini}, A. and {Saglia}, R. and {Sharples}, R.~M. and {Sternberg}, A. and {Wilman}, D. and {Wuyts}, E.},
  title = {{The Angular Momentum Distribution and Baryon Content of Star-forming Galaxies at z {\ensuremath{\sim}} 1-3}},
  journal = {\apj},
  year = {2016},
  month = aug,
  volume = {826},
  number = {2},
  eid = {214},
  pages = {214},
  doi = {10.3847/0004-637X/826/2/214},
  archiveprefix = {arXiv},
  eprint = {1510.03262},
  primaryclass = {astro-ph.GA},
  adsurl = {https://ui.adsabs.harvard.edu/abs/2016ApJ...826..214B},
}

@article{Bett2007,
  author = {{Bett}, Philip and {Eke}, Vincent and {Frenk}, Carlos S. and {Jenkins}, Adrian and {Helly}, John and {Navarro}, Julio},
  title = {{The spin and shape of dark matter haloes in the Millennium simulation of a {\ensuremath{\Lambda}} cold dark matter universe}},
  journal = {\mnras},
  year = {2007},
  month = mar,
  volume = {376},
  number = {1},
  pages = {215-232},
  doi = {10.1111/j.1365-2966.2007.11432.x},
  archiveprefix = {arXiv},
  eprint = {astro-ph/0608607},
  primaryclass = {astro-ph},
  adsurl = {https://ui.adsabs.harvard.edu/abs/2007MNRAS.376..215B},
}

@article{Toomre1964,
  author = {{Toomre}, A.},
  title = {{On the gravitational stability of a disk of stars.}},
  journal = {\apj},
  year = {1964},
  month = may,
  volume = {139},
  pages = {1217-1238},
  doi = {10.1086/147861},
  adsurl = {https://ui.adsabs.harvard.edu/abs/1964ApJ...139.1217T},
}

@article{Yoshida2003,
  author = {{Yoshida}, Naoki and {Abel}, Tom and {Hernquist}, Lars and {Sugiyama}, Naoshi},
  title = {{Simulations of Early Structure Formation: Primordial Gas Clouds}},
  journal = {\apj},
  year = {2003},
  month = aug,
  volume = {592},
  number = {2},
  pages = {645-663},
  doi = {10.1086/375810},
  archiveprefix = {arXiv},
  eprint = {astro-ph/0301645},
  primaryclass = {astro-ph},
  adsurl = {https://ui.adsabs.harvard.edu/abs/2003ApJ...592..645Y},
}

@article{Liu2020,
  author = {{Liu}, Boyuan and {Bromm}, Volker},
  title = {{Gravitational waves from Population III binary black holes formed by dynamical capture}},
  journal = {\mnras},
  year = {2020},
  month = jun,
  volume = {495},
  number = {2},
  pages = {2475-2495},
  doi = {10.1093/mnras/staa1362},
  archiveprefix = {arXiv},
  eprint = {2003.00065},
  primaryclass = {astro-ph.CO},
  adsurl = {https://ui.adsabs.harvard.edu/abs/2020MNRAS.495.2475L},
}

@article{Ostriker1999,
  author = {{Ostriker}, Eve C.},
  title = {{Dynamical Friction in a Gaseous Medium}},
  journal = {\apj},
  year = {1999},
  month = mar,
  volume = {513},
  number = {1},
  pages = {252-258},
  doi = {10.1086/306858},
  archiveprefix = {arXiv},
  eprint = {astro-ph/9810324},
  primaryclass = {astro-ph},
  adsurl = {https://ui.adsabs.harvard.edu/abs/1999ApJ...513..252O},
}

@article{Munoz-Cuartas2011,
  author = {{Mu{\~n}oz-Cuartas}, J.~C. and {Macci{\`o}}, A.~V. and {Gottl{\"o}ber}, S. and {Dutton}, A.~A.},
  title = {{The redshift evolution of {\ensuremath{\Lambda}} cold dark matter halo parameters: concentration, spin and shape}},
  journal = {\mnras},
  year = {2011},
  month = feb,
  volume = {411},
  number = {1},
  pages = {584-594},
  doi = {10.1111/j.1365-2966.2010.17704.x},
  archiveprefix = {arXiv},
  eprint = {1007.0438},
  primaryclass = {astro-ph.CO},
  adsurl = {https://ui.adsabs.harvard.edu/abs/2011MNRAS.411..584M},
}

@article{Nandal2025,
  author = {{Nandal}, Devesh and {Buldgen}, Ga{\"e}l and {Whalen}, Daniel J. and {Regan}, John and {Woods}, Tyrone E. and {Tan}, Jonathan C.},
  title = {{Rotating supermassive Pop III stars on the main sequence}},
  journal = {\aap},
  year = {2025},
  month = sep,
  volume = {701},
  eid = {A262},
  pages = {A262},
  doi = {10.1051/0004-6361/202555878},
  archiveprefix = {arXiv},
  eprint = {2506.08268},
  primaryclass = {astro-ph.SR},
  adsurl = {https://ui.adsabs.harvard.edu/abs/2025A&A...701A.262N},
}

@article{Fakhouri2010,
  author = {{Fakhouri}, Onsi and {Ma}, Chung-Pei and {Boylan-Kolchin}, Michael},
  title = {{The merger rates and mass assembly histories of dark matter haloes in the two Millennium simulations}},
  journal = {\mnras},
  year = {2010},
  month = aug,
  volume = {406},
  number = {4},
  pages = {2267-2278},
  doi = {10.1111/j.1365-2966.2010.16859.x},
  archiveprefix = {arXiv},
  eprint = {1001.2304},
  primaryclass = {astro-ph.CO},
  adsurl = {https://ui.adsabs.harvard.edu/abs/2010MNRAS.406.2267F},
}

@article{Planck2020,
  author = {{Planck Collaboration} and {Aghanim}, N. and {Akrami}, Y. and {Ashdown}, M. and {Aumont}, J. and {Baccigalupi}, C. and {Ballardini}, M. and {Banday}, A.~J. and {Barreiro}, R.~B. and {Bartolo}, N. and {Basak}, S. and {Battye}, R. and {Benabed}, K. and {Bernard}, J.-P. and {Bersanelli}, M. and {Bielewicz}, P. and {Bock}, J.~J. and {Bond}, J.~R. and {Borrill}, J. and {Bouchet}, F.~R. and {Boulanger}, F. and {Bucher}, M. and {Burigana}, C. and {Butler}, R.~C. and {Calabrese}, E. and {Cardoso}, J.-F. and {Carron}, J. and {Challinor}, A. and {Chiang}, H.~C. and {Chluba}, J. and {Colombo}, L.~P.~L. and {Combet}, C. and {Contreras}, D. and {Crill}, B.~P. and {Cuttaia}, F. and {de Bernardis}, P. and {de Zotti}, G. and {Delabrouille}, J. and {Delouis}, J.-M. and {Di Valentino}, E. and {Diego}, J.~M. and {Dor{\'e}}, O. and {Douspis}, M. and {Ducout}, A. and {Dupac}, X. and {Dusini}, S. and {Efstathiou}, G. and {Elsner}, F. and {En{\ss}lin}, T.~A. and {Eriksen}, H.~K. and {Fantaye}, Y. and {Farhang}, M. and {Fergusson}, J. and {Fernandez-Cobos}, R. and {Finelli}, F. and {Forastieri}, F. and {Frailis}, M. and {Fraisse}, A.~A. and {Franceschi}, E. and {Frolov}, A. and {Galeotta}, S. and {Galli}, S. and {Ganga}, K. and {G{\'e}nova-Santos}, R.~T. and {Gerbino}, M. and {Ghosh}, T. and {Gonz{\'a}lez-Nuevo}, J. and {G{\'o}rski}, K.~M. and {Gratton}, S. and {Gruppuso}, A. and {Gudmundsson}, J.~E. and {Hamann}, J. and {Handley}, W. and {Hansen}, F.~K. and {Herranz}, D. and {Hildebrandt}, S.~R. and {Hivon}, E. and {Huang}, Z. and {Jaffe}, A.~H. and {Jones}, W.~C. and {Karakci}, A. and {Keih{\"a}nen}, E. and {Keskitalo}, R. and {Kiiveri}, K. and {Kim}, J. and {Kisner}, T.~S. and {Knox}, L. and {Krachmalnicoff}, N. and {Kunz}, M. and {Kurki-Suonio}, H. and {Lagache}, G. and {Lamarre}, J.-M. and {Lasenby}, A. and {Lattanzi}, M. and {Lawrence}, C.~R. and {Le Jeune}, M. and {Lemos}, P. and {Lesgourgues}, J. and {Levrier}, F. and {Lewis}, A. and {Liguori}, M. and {Lilje}, P.~B. and {Lilley}, M. and {Lindholm}, V. and {L{\'o}pez-Caniego}, M. and {Lubin}, P.~M. and {Ma}, Y.-Z. and {Mac{\'\i}as-P{\'e}rez}, J.~F. and {Maggio}, G. and {Maino}, D. and {Mandolesi}, N. and {Mangilli}, A. and {Marcos-Caballero}, A. and {Maris}, M. and {Martin}, P.~G. and {Martinelli}, M. and {Mart{\'\i}nez-Gonz{\'a}lez}, E. and {Matarrese}, S. and {Mauri}, N. and {McEwen}, J.~D. and {Meinhold}, P.~R. and {Melchiorri}, A. and {Mennella}, A. and {Migliaccio}, M. and {Millea}, M. and {Mitra}, S. and {Miville-Desch{\^e}nes}, M.-A. and {Molinari}, D. and {Montier}, L. and {Morgante}, G. and {Moss}, A. and {Natoli}, P. and {N{\o}rgaard-Nielsen}, H.~U. and {Pagano}, L. and {Paoletti}, D. and {Partridge}, B. and {Patanchon}, G. and {Peiris}, H.~V. and {Perrotta}, F. and {Pettorino}, V. and {Piacentini}, F. and {Polastri}, L. and {Polenta}, G. and {Puget}, J.-L. and {Rachen}, J.~P. and {Reinecke}, M. and {Remazeilles}, M. and {Renzi}, A. and {Rocha}, G. and {Rosset}, C. and {Roudier}, G. and {Rubi{\~n}o-Mart{\'\i}n}, J.~A. and {Ruiz-Granados}, B. and {Salvati}, L. and {Sandri}, M. and {Savelainen}, M. and {Scott}, D. and {Shellard}, E.~P.~S. and {Sirignano}, C. and {Sirri}, G. and {Spencer}, L.~D. and {Sunyaev}, R. and {Suur-Uski}, A.-S. and {Tauber}, J.~A. and {Tavagnacco}, D. and {Tenti}, M. and {Toffolatti}, L. and {Tomasi}, M. and {Trombetti}, T. and {Valenziano}, L. and {Valiviita}, J. and {Van Tent}, B. and {Vibert}, L. and {Vielva}, P. and {Villa}, F. and {Vittorio}, N. and {Wandelt}, B.~D. and {Wehus}, I.~K. and {White}, M. and {White}, S.~D.~M. and {Zacchei}, A. and {Zonca}, A.},
  title = {{Planck 2018 results. VI. Cosmological parameters}},
  journal = {\aap},
  year = {2020},
  month = sep,
  volume = {641},
  eid = {A6},
  pages = {A6},
  doi = {10.1051/0004-6361/201833910},
  archiveprefix = {arXiv},
  eprint = {1807.06209},
  primaryclass = {astro-ph.CO},
  adsurl = {https://ui.adsabs.harvard.edu/abs/2020A&A...641A...6P},
}

@article{Genzel2017,
  author = {{Genzel}, R. and {F{\"o}rster Schreiber}, N.~M. and {{\"U}bler}, H. and {Lang}, P. and {Naab}, T. and {Bender}, R. and {Tacconi}, L.~J. and {Wisnioski}, E. and {Wuyts}, S. and {Alexander}, T. and {Beifiori}, A. and {Belli}, S. and {Brammer}, G. and {Burkert}, A. and {Carollo}, C.~M. and {Chan}, J. and {Davies}, R. and {Fossati}, M. and {Galametz}, A. and {Genel}, S. and {Gerhard}, O. and {Lutz}, D. and {Mendel}, J.~T. and {Momcheva}, I. and {Nelson}, E.~J. and {Renzini}, A. and {Saglia}, R. and {Sternberg}, A. and {Tacchella}, S. and {Tadaki}, K. and {Wilman}, D.},
  title = {{Strongly baryon-dominated disk galaxies at the peak of galaxy formation ten billion years ago}},
  journal = {\nat},
  year = {2017},
  month = mar,
  volume = {543},
  number = {7645},
  pages = {397-401},
  doi = {10.1038/nature21685},
  archiveprefix = {arXiv},
  eprint = {1703.04310},
  primaryclass = {astro-ph.GA},
  adsurl = {https://ui.adsabs.harvard.edu/abs/2017Natur.543..397G},
}

@article{Tacconi2013,
  author = {{Tacconi}, L.~J. and {Neri}, R. and {Genzel}, R. and {Combes}, F. and {Bolatto}, A. and {Cooper}, M.~C. and {Wuyts}, S. and {Bournaud}, F. and {Burkert}, A. and {Comerford}, J. and {Cox}, P. and {Davis}, M. and {F{\"o}rster Schreiber}, N.~M. and {Garc{\'\i}a-Burillo}, S. and {Gracia-Carpio}, J. and {Lutz}, D. and {Naab}, T. and {Newman}, S. and {Omont}, A. and {Saintonge}, A. and {Shapiro Griffin}, K. and {Shapley}, A. and {Sternberg}, A. and {Weiner}, B.},
  title = {{Phibss: Molecular Gas Content and Scaling Relations in z \raisebox{-0.5ex}\textasciitilde 1-3 Massive, Main-sequence Star-forming Galaxies}},
  journal = {\apj},
  year = {2013},
  month = may,
  volume = {768},
  number = {1},
  eid = {74},
  pages = {74},
  doi = {10.1088/0004-637X/768/1/74},
  archiveprefix = {arXiv},
  eprint = {1211.5743},
  primaryclass = {astro-ph.CO},
  adsurl = {https://ui.adsabs.harvard.edu/abs/2013ApJ...768...74T},
}

@article{Yung2024,
  author = {{Yung}, L.~Y. Aaron and {Somerville}, Rachel S. and {Nguyen}, Tri and {Behroozi}, Peter and {Modi}, Chirag and {Gardner}, Jonathan P.},
  title = {{Characterizing ultra-high-redshift dark matter halo demographics and assembly histories with the GUREFT simulations}},
  journal = {\mnras},
  year = {2024},
  month = jun,
  volume = {530},
  number = {4},
  pages = {4868-4886},
  doi = {10.1093/mnras/stae1188},
  archiveprefix = {arXiv},
  eprint = {2309.14408},
  primaryclass = {astro-ph.CO},
  adsurl = {https://ui.adsabs.harvard.edu/abs/2024MNRAS.530.4868Y},
}

@article{Abel2002,
  author = {{Abel}, Tom and {Bryan}, Greg L. and {Norman}, Michael L.},
  title = {{The Formation of the First Star in the Universe}},
  journal = {Science},
  year = {2002},
  month = jan,
  volume = {295},
  number = {5552},
  pages = {93-98},
  doi = {10.1126/science.1063991},
  archiveprefix = {arXiv},
  eprint = {astro-ph/0112088},
  primaryclass = {astro-ph},
  adsurl = {https://ui.adsabs.harvard.edu/abs/2002Sci...295...93A},
}

@article{Yoshida2006,
  author = {{Yoshida}, Naoki and {Omukai}, Kazuyuki and {Hernquist}, Lars and {Abel}, Tom},
  title = {{Formation of Primordial Stars in a {\ensuremath{\Lambda}}CDM Universe}},
  journal = {\apj},
  year = {2006},
  month = nov,
  volume = {652},
  number = {1},
  pages = {6-25},
  doi = {10.1086/507978},
  archiveprefix = {arXiv},
  eprint = {astro-ph/0606106},
  primaryclass = {astro-ph},
  adsurl = {https://ui.adsabs.harvard.edu/abs/2006ApJ...652....6Y},
}

@article{Bromm2002,
  author = {{Bromm}, Volker and {Coppi}, Paolo S. and {Larson}, Richard B.},
  title = {{The Formation of the First Stars. I. The Primordial Star-forming Cloud}},
  journal = {\apj},
  year = {2002},
  month = jan,
  volume = {564},
  number = {1},
  pages = {23-51},
  doi = {10.1086/323947},
  archiveprefix = {arXiv},
  eprint = {astro-ph/0102503},
  primaryclass = {astro-ph},
  adsurl = {https://ui.adsabs.harvard.edu/abs/2002ApJ...564...23B},
}

@article{Mayer2004,
  author = {{Mayer}, Lucio and {Quinn}, Thomas and {Wadsley}, James and {Stadel}, Joachim},
  title = {{The Evolution of Gravitationally Unstable Protoplanetary Disks: Fragmentation and Possible Giant Planet Formation}},
  journal = {\apj},
  year = {2004},
  month = jul,
  volume = {609},
  number = {2},
  pages = {1045-1064},
  doi = {10.1086/421288},
  archiveprefix = {arXiv},
  eprint = {astro-ph/0310771},
  primaryclass = {astro-ph},
  adsurl = {https://ui.adsabs.harvard.edu/abs/2004ApJ...609.1045M},
}

@inproceedings{Durisen2007,
  author = {{Durisen}, R.~H. and {Boss}, A.~P. and {Mayer}, L. and {Nelson}, A.~F. and {Quinn}, T. and {Rice}, W.~K.~M.},
  title = {{Gravitational Instabilities in Gaseous Protoplanetary Disks and Implications for Giant Planet Formation}},
  booktitle = {Protostars and Planets V},
  year = {2007},
  editor = {{Reipurth}, Bo and {Jewitt}, David and {Keil}, Klaus},
  month = jan,
  pages = {607},
  doi = {10.48550/arXiv.astro-ph/0603179},
  archiveprefix = {arXiv},
  eprint = {astro-ph/0603179},
  primaryclass = {astro-ph},
  adsurl = {https://ui.adsabs.harvard.edu/abs/2007prpl.conf..607D},
}

@article{Johnson2003,
  author = {{Johnson}, Bryan M. and {Gammie}, Charles F.},
  title = {{Nonlinear Outcome of Gravitational Instability in Disks with Realistic Cooling}},
  journal = {\apj},
  year = {2003},
  month = nov,
  volume = {597},
  number = {1},
  pages = {131-141},
  doi = {10.1086/378392},
  archiveprefix = {arXiv},
  eprint = {astro-ph/0312507},
  primaryclass = {astro-ph},
  adsurl = {https://ui.adsabs.harvard.edu/abs/2003ApJ...597..131J},
}

@article{Branchesi2023,
  author = {{Branchesi}, Marica and {Maggiore}, Michele and {Alonso}, David and {Badger}, Charles and {Banerjee}, Biswajit and {Beirnaert}, Freija and {Belgacem}, Enis and {Bhagwat}, Swetha and {Boileau}, Guillaume and {Borhanian}, Ssohrab and {Brown}, Daniel David and {Leong Chan}, Man and {Cusin}, Giulia and {Danilishin}, Stefan L. and {Degallaix}, Jerome and {De Luca}, Valerio and {Dhani}, Arnab and {Dietrich}, Tim and {Dupletsa}, Ulyana and {Foffa}, Stefano and {Franciolini}, Gabriele and {Freise}, Andreas and {Gemme}, Gianluca and {Goncharov}, Boris and {Ghosh}, Archisman and {Gulminelli}, Francesca and {Gupta}, Ish and {Kumar Gupta}, Pawan and {Harms}, Jan and {Hazra}, Nandini and {Hild}, Stefan and {Hinderer}, Tanja and {Siong Heng}, Ik and {Iacovelli}, Francesco and {Janquart}, Justin and {Janssens}, Kamiel and {Jenkins}, Alexander C. and {Kalaghatgi}, Chinmay and {Koroveshi}, Xhesika and {Li}, Tjonnie G.~F. and {Li}, Yufeng and {Loffredo}, Eleonora and {Maggio}, Elisa and {Mancarella}, Michele and {Mapelli}, Michela and {Martinovic}, Katarina and {Maselli}, Andrea and {Meyers}, Patrick and {Miller}, Andrew L. and {Mondal}, Chiranjib and {Muttoni}, Niccol{\`o} and {Narola}, Harsh and {Oertel}, Micaela and {Oganesyan}, Gor and {Pacilio}, Costantino and {Palomba}, Cristiano and {Pani}, Paolo and {Pasqualetti}, Antonio and {Perego}, Albino and {P{\'e}rigois}, Carole and {Pieroni}, Mauro and {Piccinni}, Ornella Juliana and {Puecher}, Anna and {Puppo}, Paola and {Ricciardone}, Angelo and {Riotto}, Antonio and {Ronchini}, Samuele and {Sakellariadou}, Mairi and {Samajdar}, Anuradha and {Santoliquido}, Filippo and {Sathyaprakash}, B.~S. and {Steinlechner}, Jessica and {Steinlechner}, Sebastian and {Utina}, Andrei and {Van Den Broeck}, Chris and {Zhang}, Teng},
  title = {{Science with the Einstein Telescope: a comparison of different designs}},
  journal = {\jcap},
  year = {2023},
  month = jul,
  volume = {2023},
  number = {7},
  eid = {068},
  pages = {068},
  doi = {10.1088/1475-7516/2023/07/068},
  archiveprefix = {arXiv},
  eprint = {2303.15923},
  primaryclass = {gr-qc},
  adsurl = {https://ui.adsabs.harvard.edu/abs/2023JCAP...07..068B},
}

@article{Escala2004,
  author = {{Escala}, Andr{\'e}s and {Larson}, Richard B. and {Coppi}, Paolo S. and {Mardones}, Diego},
  title = {{The Role of Gas in the Merging of Massive Black Holes in Galactic Nuclei. I. Black Hole Merging in a Spherical Gas Cloud}},
  journal = {\apj},
  year = {2004},
  month = jun,
  volume = {607},
  number = {2},
  pages = {765-777},
  doi = {10.1086/386278},
  archiveprefix = {arXiv},
  eprint = {astro-ph/0310851},
  primaryclass = {astro-ph},
  adsurl = {https://ui.adsabs.harvard.edu/abs/2004ApJ...607..765E},
}

@article{Chapon2013,
  author = {{Chapon}, Damien and {Mayer}, Lucio and {Teyssier}, Romain},
  title = {{Hydrodynamics of galaxy mergers with supermassive black holes: is there a last parsec problem?}},
  journal = {\mnras},
  year = {2013},
  month = mar,
  volume = {429},
  number = {4},
  pages = {3114-3122},
  doi = {10.1093/mnras/sts568},
  archiveprefix = {arXiv},
  eprint = {1110.6086},
  primaryclass = {astro-ph.GA},
  adsurl = {https://ui.adsabs.harvard.edu/abs/2013MNRAS.429.3114C},
}

@article{Luo2021,
  author = {{Luo}, Ziren and {Wang}, Yan and {Wu}, Yueliang and {Hu}, Wenrui and {Jin}, Gang},
  title = {{The Taiji program: A concise overview}},
  journal = {Progress of Theoretical and Experimental Physics},
  year = {2021},
  month = may,
  volume = {2021},
  number = {5},
  eid = {05A108},
  pages = {05A108},
  doi = {10.1093/ptep/ptaa083},
  adsurl = {https://ui.adsabs.harvard.edu/abs/2021PTEP.2021eA108L},
}

@article{Klessen2023,
  author = {{Klessen}, Ralf S. and {Glover}, Simon C.~O.},
  title = {{The First Stars: Formation, Properties, and Impact}},
  journal = {\araa},
  year = {2023},
  month = aug,
  volume = {61},
  pages = {65-130},
  doi = {10.1146/annurev-astro-071221-053453},
  archiveprefix = {arXiv},
  eprint = {2303.12500},
  primaryclass = {astro-ph.CO},
  adsurl = {https://ui.adsabs.harvard.edu/abs/2023ARA&A..61...65K},
}

@article{Navarro1997,
  author = {{Navarro}, Julio F. and {Frenk}, Carlos S. and {White}, Simon D.~M.},
  title = {{A Universal Density Profile from Hierarchical Clustering}},
  journal = {\apj},
  year = {1997},
  month = dec,
  volume = {490},
  number = {2},
  pages = {493-508},
  doi = {10.1086/304888},
  archiveprefix = {arXiv},
  eprint = {astro-ph/9611107},
  primaryclass = {astro-ph},
  adsurl = {https://ui.adsabs.harvard.edu/abs/1997ApJ...490..493N},
}

@article{Colpi2024,
  author = {{Colpi}, Monica and {Danzmann}, Karsten and {Hewitson}, Martin and {Holley-Bockelmann}, Kelly and {Jetzer}, Philippe and {Nelemans}, Gijs and {Petiteau}, Antoine and {Shoemaker}, David and {Sopuerta}, Carlos and {Stebbins}, Robin and {Tanvir}, Nial and {Ward}, Henry and {Weber}, William Joseph and {Thorpe}, Ira and {Daurskikh}, Anna and {Deep}, Atul and {Fern{\'a}ndez N{\'u}{\~n}ez}, Ignacio and {Garc{\'\i}a Marirrodriga}, C{\'e}sar and {Gehler}, Martin and {Halain}, Jean-Philippe and {Jennrich}, Oliver and {Lammers}, Uwe and {Larra{\~n}aga}, Jonan and {Lieser}, Maike and {L{\"u}tzgendorf}, Nora and {Martens}, Waldemar and {Mondin}, Linda and {Piris Ni{\~n}o}, Ana and {Amaro-Seoane}, Pau and {Arca Sedda}, Manuel and {Auclair}, Pierre and {Babak}, Stanislav and {Baghi}, Quentin and {Baibhav}, Vishal and {Baker}, Tessa and {Bayle}, Jean-Baptiste and {Berry}, Christopher and {Berti}, Emanuele and {Boileau}, Guillaume and {Bonetti}, Matteo and {Brito}, Richard and {Buscicchio}, Riccardo and {Calcagni}, Gianluca and {Capelo}, Pedro R. and {Caprini}, Chiara and {Caputo}, Andrea and {Castelli}, Eleonora and {Chen}, Hsin-Yu and {Chen}, Xian and {Chua}, Alvin and {Davies}, Gareth and {Derdzinski}, Andrea and {Domcke}, Valerie Fiona and {Doneva}, Daniela and {Dvorkin}, Irna and {Mar{\'\i}a Ezquiaga}, Jose and {Gair}, Jonathan and {Haiman}, Zoltan and {Harry}, Ian and {Hartwig}, Olaf and {Hees}, Aurelien and {Heffernan}, Anna and {Husa}, Sascha and {Izquierdo-Villalba}, David and {Karnesis}, Nikolaos and {Klein}, Antoine and {Korol}, Valeriya and {Korsakova}, Natalia and {Kupfer}, Thomas and {Laghi}, Danny and {Lamberts}, Astrid and {Larson}, Shane and {Le Jeune}, Maude and {Lewicki}, Marek and {Littenberg}, Tyson and {Madge}, Eric and {Mangiagli}, Alberto and {Marsat}, Sylvain and {Vilchez}, Ivan Martin and {Maselli}, Andrea and {Mathews}, Josh and {van de Meent}, Maarten and {Muratore}, Martina and {Nardini}, Germano and {Pani}, Paolo and {Peloso}, Marco and {Pieroni}, Mauro and {Pound}, Adam and {Quelquejay-Leclere}, Hippolyte and {Ricciardone}, Angelo and {Rossi}, Elena Maria and {Sartirana}, Andrea and {Savalle}, Etienne and {Sberna}, Laura and {Sesana}, Alberto and {Shoemaker}, Deirdre and {Slutsky}, Jacob and {Sotiriou}, Thomas and {Speri}, Lorenzo and {Staab}, Martin and {Steer}, Dani{\`e}le and {Tamanini}, Nicola and {Tasinato}, Gianmassimo and {Torrado}, Jesus and {Torres-Orjuela}, Alejandro and {Toubiana}, Alexandre and {Vallisneri}, Michele and {Vecchio}, Alberto and {Volonteri}, Marta and {Yagi}, Kent and {Zwick}, Lorenz},
  title = {{LISA Definition Study Report}},
  journal = {arXiv e-prints},
  year = {2024},
  month = feb,
  eid = {arXiv:2402.07571},
  pages = {arXiv:2402.07571},
  doi = {10.48550/arXiv.2402.07571},
  archiveprefix = {arXiv},
  eprint = {2402.07571},
  primaryclass = {astro-ph.CO},
  adsurl = {https://ui.adsabs.harvard.edu/abs/2024arXiv240207571C},
}

@article{Abac2026,
  author = {{Abac}, Adrian and {Abramo}, Raul and {Albanesi}, Simone and {Albertini}, Angelica and {Agapito}, Alessandro and {Agathos}, Michalis and {Albertus}, Conrado and {Andersson}, Nils and {Andrade}, Tom{\'a}s and {Andreoni}, Igor and {Angeloni}, Federico and {Antonelli}, Marco and {Antoniadis}, John and {Antonini}, Fabio and {Arca Sedda}, Manuel and {Artale}, M. Celeste and {Ascenzi}, Stefano and {Auclair}, Pierre and {Bachetti}, Matteo and {Badger}, Charles and {Banerjee}, Biswajit and {Barba-Gonz{\'a}lez}, David and {Barta}, D{\'a}niel and {Bartolo}, Nicola and {Bauswein}, Andreas and {Begnoni}, Andrea and {Beirnaert}, Freija and {Bejger}, Micha{\l} and {Belgacem}, Enis and {Bellomo}, Nicola and {Bernard}, Laura and {Bernardini}, Maria Grazia and {Bernuzzi}, Sebastiano and {Berry}, Christopher P.~L. and {Berti}, Emanuele and {Bertone}, Gianfranco and {Bettoni}, Dario and {Bezares}, Miguel and {Bhagwat}, Swetha and {Bisero}, Sofia and {Bizouard}, Marie Anne and {Blanco-Pillado}, Jose J. and {Blasi}, Simone and {Bonino}, Alice and {Borghese}, Alice and {Borghi}, Nicola and {Borhanian}, Ssohrab and {Bortolas}, Elisa and {Botticella}, Maria Teresa and {Branchesi}, Marica and {Breschi}, Matteo and {Brito}, Richard and {Brocato}, Enzo and {Broekgaarden}, Floor S. and {Bulik}, Tomasz and {Buonanno}, Alessandra and {Burgio}, Fiorella and {Burrows}, Adam and {Calcagni}, Gianluca and {Canevarolo}, Sofia and {Cappellaro}, Enrico and {Capurri}, Giulia and {Carbone}, Carmelita and {Casadio}, Roberto and {Cayuso}, Ramiro and {Cerd{\'a}-Dur{\'a}n}, Pablo and {Char}, Prasanta and {Chaty}, Sylvain and {Chiarusi}, Tommaso and {Chruslinska}, Martyna and {Cireddu}, Francesco and {Cole}, Philippa and {Colombo}, Alberto and {Colpi}, Monica and {Comp{\`e}re}, Geoffrey and {Contaldi}, Carlo and {Corman}, Maxence and {Crescimbeni}, Francesco and {Cristallo}, Sergio and {Cuoco}, Elena and {Cusin}, Giulia and {Canton}, Tito Dal and {D{\'a}lya}, Gergely and {D'Avanzo}, Paolo and {Davari}, Nazanin and {De Luca}, Valerio and {De Renzis}, Viola and {Della Valle}, Massimo and {Del Pozzo}, Walter and {De Santi}, Federico and {De Santis}, Alessio Ludovico and {Dietrich}, Tim and {Dimastrogiovanni}, Ema and {Domenech}, Guillem and {Doneva}, Daniela and {Drago}, Marco and {Dupletsa}, Ulyana and {Duval}, Hannah and {Dvorkin}, Irina and {Elias-Rosa}, Nancy and {Fairhurst}, Stephen and {Fantina}, Anthea F. and {Fasiello}, Matteo and {Fays}, Maxime and {Fender}, Rob and {Fischer}, Tobias and {Foucart}, Fran{\c{c}}ois and {Fragos}, Tassos and {Foffa}, Stefano and {Franciolini}, Gabriele and {Fumagalli}, Jacopo and {Gair}, Jonathan and {Gamba}, Rossella and {Garcia-Bellido}, Juan and {Garc{\'\i}a-Quir{\'o}s}, Cecilio and {Gergely}, L{\'a}szl{\'o} {\'A}rp{\'a}d and {Ghirlanda}, Giancarlo and {Ghosh}, Archisman and {Giacomazzo}, Bruno and {Gittins}, Fabian and {Giudice}, Ines Francesca and {Goncharov}, Boris and {Gonzalez}, Alejandra and {Gori{\'e}ly}, St{\'e}phane and {Graziani}, Luca and {Greco}, Giuseppe and {Gualtieri}, Leonardo and {Guidi}, Gianluca Maria and {Gupta}, Ish and {Haney}, Maria and {Hannam}, Mark and {Harms}, Jan and {Harutyunyan}, Arus and {Haskell}, Brynmor and {Haungs}, Andreas and {Hazra}, Nandini and {Hemming}, Gary and {Heng}, Ik Siong and {Hinderer}, Tanja and {van der Horst}, Alexander and {Hu}, Qian and {Husa}, Sascha and {Iacovelli}, Francesco and {Illuminati}, Giulia and {Inguglia}, Gianluca and {Villalba}, David Izquierdo and {Janquart}, Justin and {Janssens}, Kamiel and {Jenkins}, Alexander C. and {Jones}, Ian and {Kacskovics}, Bal{\'a}zs and {Klessen}, Ralf S. and {Kokkotas}, Kostas and {Kuan}, Hao-Jui and {Kumar}, Sumit and {Kuroyanagi}, Sachiko and {Laghi}, Danny and {Lamberts}, Astrid and {Lambiase}, Gaetano and {Larrouturou}, Fran{\c{c}}ois and {Leaci}, Paola and {Lenzi}, Michele and {Levan}, Andrew and {Li}, T.~G.~F. and {Li}, Yufeng and {Liang}, Dicong and {Limongi}, Marco and {Liu}, Boyuan and {Llanes-Estrada}, Felipe J. and {Loffredo}, Eleonora and {Long}, Oliver and {Lope-Oter}, Eva and {Lukes-Gerakopoulos}, Georgios and {Maggio}, Elisa and {Maggiore}, Michele and {Mancarella}, Michele and {Mapelli}, Michela and {Marchant}, Pablo and {Margiotta}, Annarita and {Mariotti}, Alberto and {Marriott-Best}, Alisha and {Marsat}, Sylvain and {Mart{\'\i}nez-Pinedo}, Gabriel and {Maselli}, Andrea and {Mastrogiovanni}, Simone and {Matos}, Isabela and {Melandri}, Andrea and {Mendes}, Raissa F.~P. and {de Souza}, Josiel Mendon{\c{c}}a Soares and {Mentasti}, Giorgio and {Mezcua}, Mar and {M{\"o}sta}, Philipp and {Mondal}, Chiranjib and {Moresco}, Michele and {Mukherjee}, Tista and {Muttoni}, Niccol{\`o} and {Nagar}, Alessandro and {Narola}, Harsh and {Nava}, Lara and {Moreno}, Pablo Navarro},
  title = {{The Science of the Einstein Telescope}},
  journal = {\jcap},
  year = {2026},
  month = mar,
  volume = {2026},
  number = {3},
  eid = {081},
  pages = {081},
  doi = {10.1088/1475-7516/2026/03/081},
  archiveprefix = {arXiv},
  eprint = {2503.12263},
  primaryclass = {gr-qc},
  adsurl = {https://ui.adsabs.harvard.edu/abs/2026JCAP...03..081A},
}

@article{Hall2022,
  author = {{Hall}, Evan D.},
  title = {{Cosmic Explorer: A Next-Generation Ground-Based Gravitational-Wave Observatory}},
  journal = {Galaxies},
  year = {2022},
  month = aug,
  volume = {10},
  number = {4},
  eid = {90},
  pages = {90},
  doi = {10.3390/galaxies10040090},
  adsurl = {https://ui.adsabs.harvard.edu/abs/2022Galax..10...90H},
}

@article{Luo2016,
  author = {{Luo}, Jun and {Chen}, Li-Sheng and {Duan}, Hui-Zong and {Gong}, Yun-Gui and {Hu}, Shoucun and {Ji}, Jianghui and {Liu}, Qi and {Mei}, Jianwei and {Milyukov}, Vadim and {Sazhin}, Mikhail and {Shao}, Cheng-Gang and {Toth}, Viktor T. and {Tu}, Hai-Bo and {Wang}, Yamin and {Wang}, Yan and {Yeh}, Hsien-Chi and {Zhan}, Ming-Sheng and {Zhang}, Yonghe and {Zharov}, Vladimir and {Zhou}, Ze-Bing},
  title = {{TianQin: a space-borne gravitational wave detector}},
  journal = {Classical and Quantum Gravity},
  year = {2016},
  month = feb,
  volume = {33},
  number = {3},
  eid = {035010},
  pages = {035010},
  doi = {10.1088/0264-9381/33/3/035010},
  archiveprefix = {arXiv},
  eprint = {1512.02076},
  primaryclass = {astro-ph.IM},
  adsurl = {https://ui.adsabs.harvard.edu/abs/2016CQGra..33c5010L},
}

@article{Hu2017,
  author = {{Hu}, Wen-Rui and {Wu}, Yue-Liang},
  title = {{The Taiji Program in Space for gravitational wave physics and the nature of gravity}},
  journal = {National Science Review},
  year = {2017},
  month = sep,
  volume = {4},
  number = {5},
  pages = {685-686},
  doi = {10.1093/nsr/nwx116},
  adsurl = {https://ui.adsabs.harvard.edu/abs/2017NSRev...4..685H},
}

@article{Maiolino2024,
  author = {{Maiolino}, Roberto and {Scholtz}, Jan and {Witstok}, Joris and {Carniani}, Stefano and {D'Eugenio}, Francesco and {de Graaff}, Anna and {{\"U}bler}, Hannah and {Tacchella}, Sandro and {Curtis-Lake}, Emma and {Arribas}, Santiago and {Bunker}, Andrew and {Charlot}, St{\'e}phane and {Chevallard}, Jacopo and {Curti}, Mirko and {Looser}, Tobias J. and {Maseda}, Michael V. and {Rawle}, Timothy D. and {Rodr{\'\i}guez del Pino}, Bruno and {Willott}, Chris J. and {Egami}, Eiichi and {Eisenstein}, Daniel J. and {Hainline}, Kevin N. and {Robertson}, Brant and {Williams}, Christina C. and {Willmer}, Christopher N.~A. and {Baker}, William M. and {Boyett}, Kristan and {DeCoursey}, Christa and {Fabian}, Andrew C. and {Helton}, Jakob M. and {Ji}, Zhiyuan and {Jones}, Gareth C. and {Kumari}, Nimisha and {Laporte}, Nicolas and {Nelson}, Erica J. and {Perna}, Michele and {Sandles}, Lester and {Shivaei}, Irene and {Sun}, Fengwu},
  title = {{A small and vigorous black hole in the early Universe}},
  journal = {\nat},
  year = {2024},
  month = mar,
  volume = {627},
  number = {8002},
  pages = {59-63},
  doi = {10.1038/s41586-024-07052-5},
  archiveprefix = {arXiv},
  eprint = {2305.12492},
  primaryclass = {astro-ph.GA},
  adsurl = {https://ui.adsabs.harvard.edu/abs/2024Natur.627...59M},
}

@article{Inayoshi2020,
  author = {{Inayoshi}, Kohei and {Visbal}, Eli and {Haiman}, Zolt{\'a}n},
  title = {{The Assembly of the First Massive Black Holes}},
  journal = {\araa},
  year = {2020},
  month = aug,
  volume = {58},
  pages = {27-97},
  doi = {10.1146/annurev-astro-120419-014455},
  archiveprefix = {arXiv},
  eprint = {1911.05791},
  primaryclass = {astro-ph.GA},
  adsurl = {https://ui.adsabs.harvard.edu/abs/2020ARA&A..58...27I},
}

@article{volonteri2021,
  author = {{Volonteri}, Marta and {Habouzit}, M{\'e}lanie and {Colpi}, Monica},
  title = {{The origins of massive black holes}},
  journal = {Nature Reviews Physics},
  year = {2021},
  month = sep,
  volume = {3},
  number = {11},
  pages = {732-743},
  doi = {10.1038/s42254-021-00364-9},
  archiveprefix = {arXiv},
  eprint = {2110.10175},
  primaryclass = {astro-ph.GA},
  adsurl = {https://ui.adsabs.harvard.edu/abs/2021NatRP...3..732V},
}

@article{Sesana2007,
  author = {{Sesana}, Alberto and {Volonteri}, Marta and {Haardt}, Francesco},
  title = {{The imprint of massive black hole formation models on the LISA data stream}},
  journal = {\mnras},
  year = {2007},
  month = jun,
  volume = {377},
  number = {4},
  pages = {1711-1716},
  doi = {10.1111/j.1365-2966.2007.11734.x},
  archiveprefix = {arXiv},
  eprint = {astro-ph/0701556},
  primaryclass = {astro-ph},
  adsurl = {https://ui.adsabs.harvard.edu/abs/2007MNRAS.377.1711S},
}

@article{Mehta2026,
  author = {{Mehta}, Daxal H. and {Regan}, John A. and {Prole}, Lewis},
  title = {{The growth of light seed black holes in the early Universe}},
  journal = {Nature Astronomy},
  year = {2026},
  month = jan,
  doi = {10.1038/s41550-025-02767-5},
  archiveprefix = {arXiv},
  eprint = {2601.14395},
  primaryclass = {astro-ph.GA},
  adsurl = {https://ui.adsabs.harvard.edu/abs/2026NatAs.tmp...21M},
}

@article{Smith2018,
  author = {{Smith}, Britton D. and {Regan}, John A. and {Downes}, Turlough P. and {Norman}, Michael L. and {O'Shea}, Brian W. and {Wise}, John H.},
  title = {{The growth of black holes from Population III remnants in the Renaissance simulations}},
  journal = {\mnras},
  year = {2018},
  month = nov,
  volume = {480},
  number = {3},
  pages = {3762-3773},
  doi = {10.1093/mnras/sty2103},
  archiveprefix = {arXiv},
  eprint = {1804.06477},
  primaryclass = {astro-ph.GA},
  adsurl = {https://ui.adsabs.harvard.edu/abs/2018MNRAS.480.3762S},
}

@article{Gordon2024,
  author = {{Gordon}, Simone T. and {Smith}, Britton D. and {Khochfar}, Sadegh and {Regan}, John Anthony},
  title = {{Hungry or not: how stellar-mass black holes grow (or don't) in dark matter mini-haloes at high resolution}},
  journal = {\mnras},
  year = {2024},
  month = mar,
  volume = {529},
  number = {1},
  pages = {604-627},
  doi = {10.1093/mnras/stae566},
  archiveprefix = {arXiv},
  eprint = {2401.04183},
  primaryclass = {astro-ph.GA},
  adsurl = {https://ui.adsabs.harvard.edu/abs/2024MNRAS.529..604G},
}

@article{maiolino2023b,
  author = {{Maiolino}, Roberto and {Scholtz}, Jan and {Curtis-Lake}, Emma and {Carniani}, Stefano and {Baker}, William and {de Graaff}, Anna and {Tacchella}, Sandro and {{\"U}bler}, Hannah and {D'Eugenio}, Francesco and {Witstok}, Joris and {Curti}, Mirko and {Arribas}, Santiago and {Bunker}, Andrew J. and {Charlot}, St{\'e}phane and {Chevallard}, Jacopo and {Eisenstein}, Daniel J. and {Egami}, Eiichi and {Ji}, Zhiyuan and {Jones}, Gareth C. and {Lyu}, Jianwei and {Rawle}, Tim and {Robertson}, Brant and {Rujopakarn}, Wiphu and {Perna}, Michele and {Sun}, Fengwu and {Venturi}, Giacomo and {Williams}, Christina C. and {Willott}, Chris},
  title = {{JADES. The diverse population of infant Black Holes at 4<z<11: merging, tiny, poor, but mighty}},
  journal = {arXiv e-prints},
  year = {2023},
  month = aug,
  eid = {arXiv:2308.01230},
  pages = {arXiv:2308.01230},
  doi = {10.48550/arXiv.2308.01230},
  archiveprefix = {arXiv},
  eprint = {2308.01230},
  primaryclass = {astro-ph.GA},
  adsurl = {https://ui.adsabs.harvard.edu/abs/2023arXiv230801230M},
}

@article{Harikane2023,
  author = {{Harikane}, Yuichi and {Zhang}, Yechi and {Nakajima}, Kimihiko and {Ouchi}, Masami and {Isobe}, Yuki and {Ono}, Yoshiaki and {Hatano}, Shun and {Xu}, Yi and {Umeda}, Hiroya},
  title = {{A JWST/NIRSpec First Census of Broad-line AGNs at z = 4-7: Detection of 10 Faint AGNs with M $_{BH}$ {}10$^{6}$-{}10$^{8}$ M $_{{\ensuremath{\odot}}}$ and Their Host Galaxy Properties}},
  journal = {\apj},
  year = {2023},
  month = dec,
  volume = {959},
  number = {1},
  eid = {39},
  pages = {39},
  doi = {10.3847/1538-4357/ad029e},
  archiveprefix = {arXiv},
  eprint = {2303.11946},
  primaryclass = {astro-ph.GA},
  adsurl = {https://ui.adsabs.harvard.edu/abs/2023ApJ...959...39H},
}

@article{greene2023,
  author = {{Greene}, Jenny E. and {Labbe}, Ivo and {Goulding}, Andy D. and {Furtak}, Lukas J. and {Chemerynska}, Iryna and {Kokorev}, Vasily and {Dayal}, Pratika and {Williams}, Christina C. and {Wang}, Bingjie and {Setton}, David J. and {Burgasser}, Adam J. and {Bezanson}, Rachel and {Atek}, Hakim and {Brammer}, Gabriel and {Cutler}, Sam E. and {Feldmann}, Robert and {Fujimoto}, Seiji and {Glazebrook}, Karl and {de Graaff}, Anna and {Leja}, Joel and {Marchesini}, Danilo and {Maseda}, Michael V. and {Matthee}, Jorryt and {Miller}, Tim B. and {Naidu}, Rohan P. and {Nanayakkara}, Themiya and {Oesch}, Pascal A. and {Pan}, Richard and {Papovich}, Casey and {Price}, Sedona H. and {van Dokkum}, Pieter and {Weaver}, John R. and {Whitaker}, Katherine E. and {Zitrin}, Adi},
  title = {{UNCOVER spectroscopy confirms a surprising ubiquity of AGN in red galaxies at $z>5$}},
  journal = {arXiv e-prints},
  year = {2023},
  month = sep,
  eid = {arXiv:2309.05714},
  pages = {arXiv:2309.05714},
  doi = {10.48550/arXiv.2309.05714},
  archiveprefix = {arXiv},
  eprint = {2309.05714},
  primaryclass = {astro-ph.GA},
  adsurl = {https://ui.adsabs.harvard.edu/abs/2023arXiv230905714G},
}

@article{Kocevski2025,
  author = {{Kocevski}, Dale D. and {Finkelstein}, Steven L. and {Barro}, Guillermo and {Taylor}, Anthony J. and {Calabr{\`o}}, Antonello and {Laloux}, Brivael and {Buchner}, Johannes and {Trump}, Jonathan R. and {Leung}, Gene C.~K. and {Yang}, Guang and {Dickinson}, Mark and {P{\'e}rez-Gonz{\'a}lez}, Pablo G. and {Pacucci}, Fabio and {Inayoshi}, Kohei and {Somerville}, Rachel S. and {McGrath}, Elizabeth J. and {Akins}, Hollis B. and {Bagley}, Micaela B. and {Bowler}, Rebecca A.~A. and {Bisigello}, Laura and {Carnall}, Adam and {Casey}, Caitlin M. and {Cheng}, Yingjie and {Cleri}, Nikko J. and {Costantin}, Luca and {Cullen}, Fergus and {Davis}, Kelcey and {Donnan}, Callum T. and {Dunlop}, James S. and {Ellis}, Richard S. and {Ferguson}, Henry C. and {Fujimoto}, Seiji and {Fontana}, Adriano and {Giavalisco}, Mauro and {Grazian}, Andrea and {Grogin}, Norman A. and {Hathi}, Nimish P. and {Hirschmann}, Michaela and {Huertas-Company}, Marc and {Holwerda}, Benne W. and {Illingworth}, Garth and {Juneau}, St{\'e}phanie and {Kartaltepe}, Jeyhan S. and {Koekemoer}, Anton M. and {Li}, Wenxiu and {Lucas}, Ray A. and {Magee}, Dan and {Mason}, Charlotte and {McLeod}, Derek J. and {McLure}, Ross J. and {Napolitano}, Lorenzo and {Papovich}, Casey and {Pirzkal}, Nor and {Rodighiero}, Giulia and {Santini}, Paola and {Wilkins}, Stephen M. and {Yung}, L.~Y. Aaron},
  title = {{The Rise of Faint, Red Active Galactic Nuclei at z > 4: A Sample of Little Red Dots in the JWST Extragalactic Legacy Fields}},
  journal = {\apj},
  year = {2025},
  month = jun,
  volume = {986},
  number = {2},
  eid = {126},
  pages = {126},
  doi = {10.3847/1538-4357/adbc7d},
  adsurl = {https://ui.adsabs.harvard.edu/abs/2025ApJ...986..126K},
}

@article{Kokorev2024,
  author = {{Kokorev}, Vasily and {Caputi}, Karina I. and {Greene}, Jenny E. and {Dayal}, Pratika and {Trebitsch}, Maxime and {Cutler}, Sam E. and {Fujimoto}, Seiji and {Labb{\'e}}, Ivo and {Miller}, Tim B. and {Iani}, Edoardo and {Navarro-Carrera}, Rafael and {Rinaldi}, Pierluigi},
  title = {{A Census of Photometrically Selected Little Red Dots at 4 < z < 9 in JWST Blank Fields}},
  journal = {\apj},
  year = {2024},
  month = jun,
  volume = {968},
  number = {1},
  eid = {38},
  pages = {38},
  doi = {10.3847/1538-4357/ad4265},
  archiveprefix = {arXiv},
  eprint = {2401.09981},
  primaryclass = {astro-ph.GA},
  adsurl = {https://ui.adsabs.harvard.edu/abs/2024ApJ...968...38K},
}

@article{Fan2001,
  author = {{Fan}, Xiaohui and {Narayanan}, Vijay K. and {Lupton}, Robert H. and {Strauss}, Michael A. and {Knapp}, Gillian R. and {Becker}, Robert H. and {White}, Richard L. and {Pentericci}, Laura and {Leggett}, S.~K. and {Haiman}, Zolt{\'a}n and {Gunn}, James E. and {Ivezi{\'c}}, {\v{Z}}eljko and {Schneider}, Donald P. and {Anderson}, Scott F. and {Brinkmann}, J. and {Bahcall}, Neta A. and {Connolly}, Andrew J. and {Csabai}, Istv{\'a}n and {Doi}, Mamoru and {Fukugita}, Masataka and {Geballe}, Tom and {Grebel}, Eva K. and {Harbeck}, Daniel and {Hennessy}, Gregory and {Lamb}, Don Q. and {Miknaitis}, Gajus and {Munn}, Jeffrey A. and {Nichol}, Robert and {Okamura}, Sadanori and {Pier}, Jeffrey R. and {Prada}, Francisco and {Richards}, Gordon T. and {Szalay}, Alex and {York}, Donald G.},
  title = {{A Survey of z>5.8 Quasars in the Sloan Digital Sky Survey. I. Discovery of Three New Quasars and the Spatial Density of Luminous Quasars at z\raisebox{-0.5ex}\textasciitilde6}},
  journal = {\aj},
  year = {2001},
  month = dec,
  volume = {122},
  number = {6},
  pages = {2833-2849},
  doi = {10.1086/324111},
  archiveprefix = {arXiv},
  eprint = {astro-ph/0108063},
  primaryclass = {astro-ph},
  adsurl = {https://ui.adsabs.harvard.edu/abs/2001AJ....122.2833F},
}

@article{fan2006,
  author = {{Fan}, Xiaohui and {Strauss}, Michael A. and {Richards}, Gordon T. and {Hennawi}, Joseph F. and {Becker}, Robert H. and {White}, Richard L. and {Diamond-Stanic}, Aleksandar M. and {Donley}, Jennifer L. and {Jiang}, Linhua and {Kim}, J. Serena and {Vestergaard}, Marianne and {Young}, Jason E. and {Gunn}, James E. and {Lupton}, Robert H. and {Knapp}, Gillian R. and {Schneider}, Donald P. and {Brandt}, W.~N. and {Bahcall}, Neta A. and {Barentine}, J.~C. and {Brinkmann}, J. and {Brewington}, Howard J. and {Fukugita}, Masataka and {Harvanek}, Michael and {Kleinman}, S.~J. and {Krzesinski}, Jurek and {Long}, Dan and {Neilsen}, Eric H., Jr. and {Nitta}, Atsuko and {Snedden}, Stephanie A. and {Voges}, Wolfgang},
  title = {{A Survey of z>5.7 Quasars in the Sloan Digital Sky Survey. IV. Discovery of Seven Additional Quasars}},
  journal = {\aj},
  year = {2006},
  month = mar,
  volume = {131},
  number = {3},
  pages = {1203-1209},
  doi = {10.1086/500296},
  archiveprefix = {arXiv},
  eprint = {astro-ph/0512080},
  primaryclass = {astro-ph},
  adsurl = {https://ui.adsabs.harvard.edu/abs/2006AJ....131.1203F},
}

@article{Banados2016,
  author = {{Ba{\~n}ados}, E. and {Venemans}, B.~P. and {Decarli}, R. and {Farina}, E.~P. and {Mazzucchelli}, C. and {Walter}, F. and {Fan}, X. and {Stern}, D. and {Schlafly}, E. and {Chambers}, K.~C. and {Rix}, H.-W. and {Jiang}, L. and {McGreer}, I. and {Simcoe}, R. and {Wang}, F. and {Yang}, J. and {Morganson}, E. and {De Rosa}, G. and {Greiner}, J. and {Balokovi{\'c}}, M. and {Burgett}, W.~S. and {Cooper}, T. and {Draper}, P.~W. and {Flewelling}, H. and {Hodapp}, K.~W. and {Jun}, H.~D. and {Kaiser}, N. and {Kudritzki}, R.-P. and {Magnier}, E.~A. and {Metcalfe}, N. and {Miller}, D. and {Schindler}, J.-T. and {Tonry}, J.~L. and {Wainscoat}, R.~J. and {Waters}, C. and {Yang}, Q.},
  title = {{The Pan-STARRS1 Distant z > 5.6 Quasar Survey: More than 100 Quasars within the First Gyr of the Universe}},
  journal = {\apjs},
  year = {2016},
  month = nov,
  volume = {227},
  number = {1},
  eid = {11},
  pages = {11},
  doi = {10.3847/0067-0049/227/1/11},
  archiveprefix = {arXiv},
  eprint = {1608.03279},
  primaryclass = {astro-ph.GA},
  adsurl = {https://ui.adsabs.harvard.edu/abs/2016ApJS..227...11B},
}

@article{Taylor2026,
  author = {{Taylor}, Anthony J. and {Finkelstein}, Steven L. and {Akins}, Hollis B. and {Kokorev}, Vasily and {Arrabal Haro}, Pablo and {Koudmani}, Sophie and {Casey}, Caitlin M. and {Dickinson}, Mark and {Jeon}, Junehyoung and {Brooks}, Madisyn and {Bromm}, Volker and {Conselice}, Christopher J. and {Davis}, Kelcey and {Donnan}, Callum T. and {Faisst}, Andreas L. and {Fujimoto}, Seiji and {Furtak}, Lukas J. and {Giavalisco}, Mauro and {Gupta}, Ansh R. and {Hirschmann}, Michaela and {Hsiao}, Tiger Yu-Yang and {Kocevski}, Dale D. and {Koekemoer}, Anton M. and {Leung}, Gene C.~K. and {Lucas}, Ray A. and {Napolitano}, Lorenzo and {Papovich}, Casey and {Pentericci}, Laura and {Perez-Gonzalez}, Pablo G. and {Somerville}, Rachel S. and {Walter}, Fabian and {Yung}, L.~Y. Aaron},
  title = {{Little Red Dots with Broad H$β$ Lines Reveal the Black Hole Mass Function at 7<z<9.5}},
  journal = {arXiv e-prints},
  year = {2026},
  month = oct,
  eid = {arXiv:2610.37564},
  pages = {arXiv:2610.37564},
  doi = {10.48550/arXiv.2610.37564},
  archiveprefix = {arXiv},
  eprint = {2610.37564},
  primaryclass = {astro-ph.GA},
  adsurl = {https://ui.adsabs.harvard.edu/abs/2026arXiv261037564T},
}

@article{Salpeter1955,
  author = {{Salpeter}, Edwin E.},
  title = {{The Luminosity Function and Stellar Evolution.}},
  journal = {\apj},
  year = {1955},
  month = jan,
  volume = {121},
  pages = {161},
  doi = {10.1086/145971},
  adsurl = {https://ui.adsabs.harvard.edu/abs/1955ApJ...121..161S},
}

@article{Bryan1998,
  author = {{Bryan}, Greg L. and {Norman}, Michael L.},
  title = {{Statistical Properties of X-Ray Clusters: Analytic and Numerical Comparisons}},
  journal = {\apj},
  year = {1998},
  month = mar,
  volume = {495},
  number = {1},
  pages = {80-99},
  doi = {10.1086/305262},
  archiveprefix = {arXiv},
  eprint = {astro-ph/9710107},
  primaryclass = {astro-ph},
  adsurl = {https://ui.adsabs.harvard.edu/abs/1998ApJ...495...80B},
}

@article{Rusta2026,
  author = {{Rusta}, Elka and {Salvadori}, Stefania and {Maiolino}, Roberto and {Gelli}, Viola and {Koutsouridou}, Ioanna and {Carniani}, Stefano and {{\"U}bler}, Hannah and {Marconi}, Alessandro and {Schaerer}, Daniel},
  title = {{The Pristine He II Emitter near GN-z11: Constraining the Mass Distribution of the First Stars}},
  journal = {\apjl},
  year = {2026},
  month = may,
  volume = {1003},
  number = {1},
  eid = {L14},
  pages = {L14},
  doi = {10.3847/2041-8213/ae64e1},
  archiveprefix = {arXiv},
  eprint = {2603.20363},
  primaryclass = {astro-ph.GA},
  adsurl = {https://ui.adsabs.harvard.edu/abs/2026ApJ..1003L..14R},
}

@article{Bardeen1972,
  author = {{Bardeen}, James M. and {Press}, William H. and {Teukolsky}, Saul A.},
  title = {{Rotating Black Holes: Locally Nonrotating Frames, Energy Extraction, and Scalar Synchrotron Radiation}},
  journal = {\apj},
  year = {1972},
  month = dec,
  volume = {178},
  pages = {347-370},
  doi = {10.1086/151796},
  adsurl = {https://ui.adsabs.harvard.edu/abs/1972ApJ...178..347B},
}

@article{Chandrasekhar1943,
  author = {{Chandrasekhar}, S.},
  title = {{Dynamical Friction. I. General Considerations: the Coefficient of Dynamical Friction.}},
  journal = {\apj},
  year = {1943},
  month = mar,
  volume = {97},
  pages = {255},
  doi = {10.1086/144517},
  adsurl = {https://ui.adsabs.harvard.edu/abs/1943ApJ....97..255C},
}

@article{Mei2021,
  author = {{Mei}, Jianwei and {Bai}, Yan-Zheng and {Bao}, Jiahui and {Barausse}, Enrico and {Cai}, Lin and {Canuto}, Enrico and {Cao}, Bin and {Chen}, Wei-Ming and {Chen}, Yu and {Ding}, Yan-Wei and {Duan}, Hui-Zong and {Fan}, Huimin and {Feng}, Wen-Fan and {Fu}, Honglin and {Gao}, Qing and {Gao}, TianQuan and {Gong}, Yungui and {Gou}, Xingyu and {Gu}, Chao-Zheng and {Gu}, De-Feng and {He}, Zi-Qi and {Hendry}, Martin and {Hong}, Wei and {Hu}, Xin-Chun and {Hu}, Yi-Ming and {Hu}, Yuexin and {Huang}, Shun-Jia and {Huang}, Xiang-Qing and {Jiang}, Qinghua and {Jiang}, Yuan-Ze and {Jiang}, Yun and {Jiang}, Zhen and {Jin}, Hong-Ming and {Korol}, Valeriya and {Li}, Hong-Yin and {Li}, Ming and {Li}, Ming and {Li}, Pengcheng and {Li}, Rongwang and {Li}, Yuqiang and {Li}, Zhu and {Li}, Zhulian and {Li}, Zhu-Xi and {Liang}, Yu-Rong and {Liang}, Zheng-Cheng and {Liao}, Fang-Jie and {Liu}, Qi and {Liu}, Shuai and {Liu}, Yan-Chong and {Liu}, Li and {Liu}, Pei-Bo and {Liu}, Xuhui and {Liu}, Yuan and {Lu}, Xiong-Fei and {Lu}, Yang and {Lu}, Ze-Huang and {Luo}, Yan and {Luo}, Zhi-Cai and {Milyukov}, Vadim and {Ming}, Min and {Pi}, Xiaoyu and {Qin}, Chenggang and {Qu}, Shao-Bo and {Sesana}, Alberto and {Shao}, Chenggang and {Shi}, Changfu and {Su}, Wei and {Tan}, Ding-Yin and {Tan}, Yujie and {Tan}, Zhuangbin and {Tu}, Liang-Cheng and {Wang}, Bin and {Wang}, Cheng-Rui and {Wang}, Fengbin and {Wang}, Guan-Fang and {Wang}, Haitian and {Wang}, Jian and {Wang}, Lijiao and {Wang}, Panpan and {Wang}, Xudong and {Wang}, Yan and {Wang}, Yi-Fan and {Wei}, Ran and {Wu}, Shu-Chao and {Xiao}, Chun-Yu and {Xu}, Xiao-Shi and {Xue}, Chao and {Yang}, Fang-Chao and {Yang}, Liang and {Yang}, Ming-Lin and {Yang}, Shan-Qing and {Ye}, Bobing and {Yeh}, Hsien-Chi and {Yu}, Shenghua and {Zhai}, Dongsheng and {Zhang}, Caishi and {Zhang}, Haitao and {Zhang}, Jian-dong and {Zhang}, Jie and {Zhang}, Lihua and {Zhang}, Xin and {Zhang}, Xuefeng and {Zhou}, Hao and {Zhou}, Ming-Yue and {Zhou}, Ze-Bing and {Zhu}, Dong-Dong and {Zi}, Tie-Guang and {Luo}, Jun},
  title = {{The TianQin project: Current progress on science and technology}},
  journal = {Progress of Theoretical and Experimental Physics},
  year = {2021},
  month = may,
  volume = {2021},
  number = {5},
  eid = {05A107},
  pages = {05A107},
  doi = {10.1093/ptep/ptaa114},
  archiveprefix = {arXiv},
  eprint = {2008.10332},
  primaryclass = {gr-qc},
  adsurl = {https://ui.adsabs.harvard.edu/abs/2021PTEP.2021eA107M},
}

@article{Boley2010,
  author = {{Boley}, Aaron C. and {Hayfield}, Tristen and {Mayer}, Lucio and {Durisen}, Richard H.},
  title = {{Clumps in the outer disk by disk instability: Why they are initially gas giants and the legacy of disruption}},
  journal = {\icarus},
  year = {2010},
  month = jun,
  volume = {207},
  number = {2},
  pages = {509-516},
  doi = {10.1016/j.icarus.2010.01.015},
  archiveprefix = {arXiv},
  eprint = {0909.4543},
  primaryclass = {astro-ph.EP},
  adsurl = {https://ui.adsabs.harvard.edu/abs/2010Icar..207..509B},
}

@article{Tamburello2015,
  author = {{Tamburello}, Valentina and {Mayer}, Lucio and {Shen}, Sijing and {Wadsley}, James},
  title = {{A lower fragmentation mass scale in high-redshift galaxies and its implications on giant clumps: a systematic numerical study}},
  journal = {\mnras},
  year = {2015},
  month = nov,
  volume = {453},
  number = {3},
  pages = {2490-2514},
  doi = {10.1093/mnras/stv1695},
  archiveprefix = {arXiv},
  eprint = {1412.3319},
  primaryclass = {astro-ph.GA},
  adsurl = {https://ui.adsabs.harvard.edu/abs/2015MNRAS.453.2490T},
}

@article{Mayer2025,
  author = {{Mayer}, Lucio and {van Donkelaar}, Floor and {Messa}, Matteo and {Capelo}, Pedro R. and {Adamo}, Angela},
  title = {{In Situ Formation of Star Clusters at z > 7 via Galactic Disk Fragmentation: Shedding Light on Ultracompact Clusters and Overmassive Black Holes Seen by JWST}},
  journal = {\apjl},
  year = {2025},
  month = mar,
  volume = {981},
  number = {2},
  eid = {L28},
  pages = {L28},
  doi = {10.3847/2041-8213/adadfe},
  archiveprefix = {arXiv},
  eprint = {2411.00670},
  primaryclass = {astro-ph.GA},
  adsurl = {https://ui.adsabs.harvard.edu/abs/2025ApJ...981L..28M},
}

@article{Haemmerle2020,
  author = {{Haemmerl{\'e}}, L. and {Mayer}, L. and {Klessen}, R.~S. and {Hosokawa}, T. and {Madau}, P. and {Bromm}, V.},
  title = {{Formation of the First Stars and Black Holes}},
  journal = {\ssr},
  year = {2020},
  month = apr,
  volume = {216},
  number = {4},
  eid = {48},
  pages = {48},
  doi = {10.1007/s11214-020-00673-y},
  archiveprefix = {arXiv},
  eprint = {2003.10533},
  primaryclass = {astro-ph.GA},
  adsurl = {https://ui.adsabs.harvard.edu/abs/2020SSRv..216...48H},
}

@article{Clark2011b,
  author = {{Clark}, Paul C. and {Glover}, Simon C.~O. and {Smith}, Rowan J. and {Greif}, Thomas H. and {Klessen}, Ralf S. and {Bromm}, Volker},
  title = {{The Formation and Fragmentation of Disks Around Primordial Protostars}},
  journal = {Science},
  year = {2011},
  month = feb,
  volume = {331},
  number = {6020},
  pages = {1040},
  doi = {10.1126/science.1198027},
  archiveprefix = {arXiv},
  eprint = {1101.5284},
  primaryclass = {astro-ph.CO},
  adsurl = {https://ui.adsabs.harvard.edu/abs/2011Sci...331.1040C},
}
\bibliographystyle{aasjournalv7}

\end{document}